\documentclass[twocolumn,showpacs,showkeys,amssymb,nobibnotes,superscriptaddress,aps,prd,bibnotes]{revtex4-2}

\usepackage{graphicx}
\usepackage{dcolumn}
\usepackage{bm}
\usepackage[T1]{fontenc}
\usepackage{mathptmx}
\usepackage{mathrsfs}
\usepackage{physics}
\usepackage{simplewick}
\usepackage{amsthm,amsmath,amssymb}
\usepackage[bottom]{footmisc}

\usepackage{amssymb}
\usepackage{amsmath}
\usepackage{amsfonts}
\usepackage{bm}
\usepackage{graphicx}
\usepackage{subfigure}
\usepackage[dvipsnames]{xcolor}
\usepackage{calc}
\usepackage{epsfig}
\usepackage{color}
\usepackage[colorlinks,citecolor=blue]{hyperref}
\begin{document}
\bibliographystyle{apsrev4-1}
\newcommand{\be}{\begin{equation}}
\newcommand{\ee}{\end{equation}}
\newcommand{\bs}{\begin{split}}
\newcommand{\es}{\end{split}}
\newcommand{\R}[1]{\textcolor{red}{#1}}
\newcommand{\B}[1]{\textcolor{blue}{#1}}

\title{Distinguishing Quantum and Classical Gravity via Non-Stationary Test Mass Dynamics}

\author{Wenjie Zhong}
\affiliation{National Gravitation Laboratory, MOE Key Laboratory of Fundamental Physical Quantities Measurement, Hubei Key Laboratory of Gravitation and Quantum Physics, School of Physics, Huazhong University of Science and Technology, Wuhan 430074, China}
\author{Yubao Liu}
\affiliation{National Gravitation Laboratory, MOE Key Laboratory of Fundamental Physical Quantities Measurement, Hubei Key Laboratory of Gravitation and Quantum Physics, School of Physics, Huazhong University of Science and Technology, Wuhan 430074, China}
\author{Yiqiu Ma}
\email{myqphy@hust.edu.cn}
\affiliation{National Gravitation Laboratory, MOE Key Laboratory of Fundamental Physical Quantities Measurement, Hubei Key Laboratory of Gravitation and Quantum Physics, School of Physics, Huazhong University of Science and Technology, Wuhan 430074, China}

\date{\today}

\begin{abstract}
     Classical gravity theory predicts a state-dependent gravitational potential for a quantum test mass, leading to nonlinear Schr{\"o}dinger-Newton\,(SN) state evolution that contrasts with quantum gravity.  Testing the effect of SN evolution can provide evidence for distinguishing quantum gravity and classical gravity, which is challenging to realize in the stationary optomechanical systems as analyzed in previous works [Phys.\,Rev.\,D 107, 024004 (2023), Phys.\,Rev.\,D 111, 062004 (2025)]. This work is devoted to analyzing the possibility of capturing the signature of SN theory during the non-stationary evolution of the test mass under the optomechanical measurement, where the second-order moments of a test mass can exhibit a distinctive oscillatory behavior. We show that this feature manifest in the non-stationary noise spectrum of outgoing light as additional peaks structures, although resolving these structures in practical experiments requires a larger number of repetitive trials with our sampling parameters, which is cost-prohibitive. To address this issue, we further employ statistical inference methods to extract more comprehensive information, thereby reducing the required number of experimental repetitions. Through Mock-Data simulations, we demonstrate that only 10 experimental trials of 40 seconds each are sufficient to reduce the false alarm rate for distinguishing between the two models to below one percent.
\end{abstract}

\maketitle

\section{Introduction}\label{section:introduction}
Testing the quantum nature of the gravitational field by designing experiments to distinguish quantum gravity and classical gravity is an important direction of experimental gravitational physics\,\cite{RevModPhys.97.015003,Gan2016,PhysRevLett.119.240401,PhysRevLett.119.240402,Miao2020,Helou2017,Datta_2021,PhysRevX.13.041040,PRXQuantum.2.030330}.
Semi-classical gravity, where the quantum expectation of matter distribution sources a classical spacetime, predicts a gravity potential that depends on the quantum state of the matter\,\cite{Mueller1962,Rosenfeld1963,Hu_Verdaguer_2020}, which has different phenomenological consequences from that of quantum gravity.  For instance, under the influence of this state-dependent gravity, the quantum evolution of the source matter's quantum state $|\psi\rangle$ satisfies the Schrödinger-Newton\,(SN) equation\,\cite{Diosi1989,Penrose1996,Diosi1998,Carlip_2008,Bahrami_2014,Adler2016}
\begin{equation}
\begin{split}
 &i\hbar\frac{\partial|\psi(t)\rangle}{\partial t}=(\hat{H}_0+\hat{V}_{\rm{SN}}(|\psi(t)\rangle))|\psi(t)\rangle,\\
 &\qquad\hat{V}_{\rm{SN}}=-GM^2\int dx'\frac{|\langle x'|\psi(t)\rangle|^2}{|x'-\hat{x}|},
\end{split}
\end{equation}
where $\hat{V}_{\rm{SN}}$ is the SN self-gravity potential. Quantum states evolved under this SN equation have a nontrivial nonlinear feature, contrasting the quantum state evolution in standard quantum mechanics\,(QM). Moreover, this nonlinear quantum state evolution offers an opportunity to test the SN theory, hence distinguishing quantum gravity theory and semi-classical gravity theory.

Applying SN theory to the macroscopic test mass mirror, the SN self-gravity potential can be further simplified to a quadratic form\,\cite {Yang2013}:
\begin{equation}\label{eq:Hamiltonian of SN gravity}
    \hat{H}_{\rm{SN}}=\frac{1}{2}M\omega_{\rm{SN}}^2(\hat{x}-\langle \hat{x}\rangle)^2,
\end{equation}
which is obtained when the uncertainty of center-of-mass position is much less than the internal atomic fluctuation of the crystal lattice.  The SN frequency is $\omega_{\rm{SN}}=\sqrt{Gm/(6\sqrt{\pi}\Delta x_{\rm int}^3)}$\,\cite{Giulini_2011,Yang2013,Giulini_2014}, where $m$ is the mass of atom and $\Delta x_{\rm int}$ is uncertainty of internal atomic fluctuations. The evolution of the test mass quantum state can be probed by coupling the test mass with the optical field, i.e., an optomechanical system\,\cite{Aspelmyer2012,Schnabel2015,PhysRevLett.124.221102,PhysRevA.107.032410,Yu2020,Mason2019,Rossi2018,PhysRevLett.105.161101,Yu2020,Aggarwal2020,10.1063/5.0118606,Smetana_2024,Chua2023}. The test mass motion information flows into the optical field quadrature and gets measured. In contrast, the self-gravity effect predicted by the quantum gravity theory at the quadratic order of mirror displacement is zero, which means the evolution of a test mass oscillator under the quantum self-gravity is the same as that of standard quantum mechanics.

Interaction between the optical field and the test mass mirror will generate optomechanical entanglement, while continuously measuring the optical field quadrature will collapse the joint optomechanical entangled state onto the conditional mechanical state, which generates a quantum trajectory\,\cite{Chen_2013,Rossi2019}. In this case, the conditional expectation value of the mechanical displacement $\langle\hat x\rangle$ becomes a conditional displacement $\langle\psi_c|\hat x|\psi_c\rangle$, and sources the classical gravity. The SN theory, incorporating the influence of quantum measurement in the above-mentioned way, is named as causal-conditional SN theory\,(CCSN) and has been systematically developed in\,\cite{Liu2023,Liu2025,Miki2025}. Though there are other approaches to treat the influence of quantum measurement\,(for instance, see\,\cite{Helou2017}), the causal conditional formalism best fits our intuition of continuous quantum measurement.

Previous works about the CCSN phenomenology focus on measuring the power spectral density of the outgoing optical quadratures\,(or their correlations) after the system rings down to the stationary steady states\,\cite{Liu2023,Liu2025,Miki2025}. The results show the same power spectral shape as in QG\,(or equivalently, standard QM), while its magnitude is slightly different. Distinguishing such a difference requires demanding experimental conditions and a detailed knowledge of the noise budget for calibration. Moreover,  Miki\,\emph{et al}.\,\cite{Miki2025} argued that such a difficulty could be circumvented by introducing the delayed-measurement, or the non-stationary measurement schemes, the former of which has been carefully analyzed in\,\cite{Miki2025}. This work targets analyzing the latter non-stationary scheme in detail, aiming to explore distinct features in the non-stationary evolution of the test mass under the influence of semi-classical self-gravity and the quantum measurement process.

In this work, we first investigate the dynamics of the test mass under the CCSN framework during its non-stationary evolution phase. Our analysis shows that the conditional second-order moments of the test mass exhibit distinctive oscillatory behavior under specific conditions—a phenomenon that can be distinct from QG. We identify three important conditions to maintain this oscillatory behavior: weak measurement strength, high SN frequency, and an appropriately initial Gaussian state. Under these conditions, we employ the Wigner-Ville\,(WV) spectrum to extract the characteristic frequencies from the non-stationary evolution of the second-order moments. These frequencies manifest as additional peaks in the WV spectrum that are unique to CCSN theory. 

Furthermore, we develop an efficient statistical inference framework that significantly reduces the number of experimental repetitions. Our results show that only 10 experimental repetitions with $40$-second duration each are sufficient to distinguish between the two models with error rates below 1\%, providing a possible practical pathway for testing the quantum nature of gravitational field.

We organize the paper structure as follows: Sec.\,\ref{section:optomechanical system configuration} briefly reviews the result about CCSN theory in a steady optomechanical system. Sec.\,\ref{section:Riccati equation} uses the numerical method to analyze the non-stationary dynamical evolution of the test mass's second-order moments, while Sec.\,\ref{section:non-stational spectrum} uses the non-stationary spectrum theory to demonstrate the distinct signatures of non-stationary SN evolution. Furthermore, a mock data analysis is performed in Sec.\,\ref{section:statistical inference}, where a series of simulated data is generated and analyzed using the statistical inference method. Finally, Section\,\ref{section:summary and discussion} presents the summary and conclusion.

\section{Review of the Stationary CCSN optomechanics}\label{section:optomechanical system configuration}
In this section, we review the previous result of the CCSN optomechanics, where the physical system is introduced with the necessary background of CCSN theory in the Schr{\"o}dinger picture\,\cite{Liu2023}, and the stationary spectrum of the outgoing optical quadrature is presented. The material in this section will be useful in analyzing the non-stationary CCSN optomechanics.

\subsection{Optomechanical system under SN self-gravity}
In SN theory, the Hamiltonian of an optomechanical system can be expressed as:
\begin{equation}\label{eq: Full formula of Hamiltonian of optomechanical system}
\begin{split}
\hat{H}_{\rm{tot}} =&\frac{\hat{p}^2}{2M}+\frac{1}{2}M\omega_m^2\hat{x}^2+\frac{1}{2}M\omega_{\rm{SN}}^2(\hat{x}-\langle \hat{x}\rangle)^2-\hbar\alpha\hat{a}_1\hat{x}\\
&+i\hbar\sqrt{2\gamma}\int \frac{d\omega}{2\pi} \left(\hat{a}^\dagger\hat{c}_{\omega+\omega_0}-\hat{a}\hat{c}^\dagger_{\omega+\omega_0}\right),\\
\end{split}
\end{equation}
where $\hat{x}$ and $\hat{p}$ are the position and momentum operators of the test mass; $\hat{a}$ and $\hat{a}^\dagger$ are the annihilation and creation operators of the cavity mode; $\hat{c}$ and $\hat{c}^\dagger$ are the annihilation and creation operators of the external continuous optical field; $\omega_m$ is the mechanical frequency of the test mass, $\gamma$ is the decay rate of the cavity mode and $\alpha=2G/\sqrt{\gamma}$ is the optomechanical coupling strength with $G = \sqrt{P_{\mathrm{cav}} \omega_0 / (\hbar L c)}$\,($L$ is the cavity length).  In this work, we assume the bad cavity approximation\,($\omega_m \ll \gamma$). The amplitude $\hat{a}_1$ and phase quadratures $\hat{a}_2$ are defined by employing the two-photon formalism\,\cite{Chen_2013}:
\begin{equation}
    \hat{a}_1=\frac{1}{\sqrt{2}}(\hat{c}+\hat{c}^\dagger),\quad \hat{a}_2=\frac{1}{\sqrt{2}i}(\hat{c}-\hat{c}^\dagger).
\end{equation}

\subsection{CCSN optomechanics}
In short, the causal conditional SN theory means that the $\langle \hat x\rangle$ in the SN term should be the conditional mean displacement $\langle \hat x\rangle_c$, driven by the optical measurement data. Essentially, it is the measurement-induced conditional mean displacement that sources the self-gravity. The evolution of $\langle \hat x\rangle_c$ is conditional on the measurement data of the outgoing optical phase quadratures:
\begin{equation}\label{eq:measurement result}
    \tilde{y}(t)=\alpha\langle\hat{x}(t)\rangle_c+dW(t)/(\sqrt{2}dt).
\end{equation}
In\,\cite{Liu2023}, the stochastic master equation\,(SME) of the conditional test mass state in CCSN is derived as:
\begin{equation}\label{eq:SME}
    \begin{split}
            \frac{d\hat{\rho} }{dt}=& -\frac{i}{\hbar}[\hat{H}_{\rm{0}},\hat{\rho}]+\frac{\alpha}{\sqrt{2}}\{\hat{x}-\langle\hat{x}\rangle,\hat{\rho}\}\frac{dW}{dt}-\frac{\alpha^2}{4}[\hat{x},[\hat{x},\hat{\rho}]],
    \end{split}
\end{equation}
where the $\hat{\rho}$ is the conditional density matrix of the test mass, and $\hat{H}_0$ is the test mass part of the Hamiltonian. The first term describes the free evolution of the system in SN theory, the second and third terms correspond to the stochastic and Lindblad terms induced by the optical sensing. The evolution of the conditional statistical moments of the mechanical motion can be derived from the SME, where the first-order moments satisfy:
\setlength{\jot}{10pt} 
\begin{align}
    \frac{d\langle \hat{x}\rangle_c}{dt} =& \frac{\langle \hat{p}\rangle_c}{M} + \sqrt{2}\alpha V_{xx}(t)\frac{dW}{dt}, \label{eq:the evolultion of the x(classical noise)}\\
    \frac{d\langle \hat{p}\rangle_c}{dt} = &-M \omega_m^2 \langle \hat{x}\rangle_c - \gamma_m\langle \hat{p} \rangle_c+ \sqrt{2}\alpha V_{xp}(t)\frac{dW}{dt} \label{eq:the evolultion of the p(classical noise)}\\ 
    &+\sqrt{\hbar M\omega_m\gamma_m\coth\frac{\hbar\omega_m}{2k_{b}T}}\frac{dW_n}{dt}\notag,
\end{align}
where the parameter $\gamma_m$ is the mechanical damping rate, and $dW_n$ is the Wiener increment caused by thermal noise, which is uncorrelated with $dW$. The empirical introduction of the decay term $-\gamma_m\langle\hat{p}\rangle$ and the Brownian motion terms arises from the classical prescription of thermal noise and the fluctuation-dissipation theorem, which are carefully discussed in\,\cite{Liu2023,Liu2025, Helou2017}.
Note that in semi-classical gravity, there exist different prescriptions for treating thermal noise\,(for details, see\,\cite{Helou2017}). In the main text, we will follow the above classical prescriptions while leaving the quantum prescriptions in the appendix. For quantum thermal noise prescription, while the main features remain unchanged, this approach imposes slightly less strict requirements on experimental parameters.

Besides, we can also obtain the Riccati equations of the evolution of the conditional second-order moments:
\begin{align}
    \dot{V}_{xx} &= \frac{2}{M}V_{xp} - 2\alpha^2V_{xx}^2, \label{eq:Vxx_classical} \\
    \dot{V}_{xp} &= \frac{V_{pp}}{M} - M\omega_q^2V_{xx} - 2\alpha^2V_{xp}V_{xx},\label{eq:Vxp_classical}  \\
    \dot{V}_{pp} &= -2M\omega_q^2V_{xp} - 2\alpha^2V_{xp}^2 + \frac{\hbar^2\alpha^2}{2}. \label{eq:Vpp_classical}
\end{align}
Note that there is no thermal noise term in the equations of second-order moments due to the classical prescription of thermal noise\,\cite{Liu2025,Helou2017}.

The conditional displacement of the test mass can be obtained by solving differential equation Eq.~\eqref{eq:the evolultion of the x(classical noise)} and Eq.~\eqref{eq:the evolultion of the p(classical noise)}:
\begin{equation}\label{eq:evolution of x(classical noise)}
    \langle \hat{x}(t)\rangle_c = e^{-\frac{\gamma_m}{2}t}\left[x^{(0)}(t)+x^m(t)+x^{\rm th}(t)\right],
\end{equation}
where $x^{(0)}(t)$, $x^m(t)$ and $x^{\rm th}(t)$ respectively are
\begin{align}
    x^{(0)}(t) &= x_0\cos(\omega_{mc} t)
    + \frac{2p_0 + Mx_0\gamma_m}{2M\omega_{mc}}\sin(\omega_{mc} t), \\
    x^m(t) &= \sqrt{2}\alpha \int^t_0 ds\ e^{\frac{\gamma_m}{2}s} \bigg[ 
    V_{xx}(s)\cos{\omega_{mc} (s-t)} \notag \\
    &\quad - \left(\frac{V_{xp}(s)}{M\omega_{mc}} 
    + \frac{\gamma_m}{2\omega_{mc}}V_{xx}(s)\right)\sin{\omega_{mc}(s-t)} \bigg] \frac{dW}{ds}, \label{eq:measurement of x}\\
    x^{\rm th}(t) &= \sqrt{\frac{\hbar \omega_m\gamma_m}{M\omega_{mc}^2}\coth\frac{\hbar\omega_m}{2k_{b}T}} 
    \int^t_0 ds\ e^{\frac{\gamma_m}{2}s} \sin{\omega_{mc}(t-s)}\frac{dW_n}{ds},\label{eq:measurement of th}
\end{align}
where the $x_0$ and $p_0$ are the quantum expectation values of the initial state of the test mass. The term $x^{(0)}(t)$ describes the free motion of the test mass's  initial state, $x^m(t)$ accounts for the influence of the measurement process on the test mass, and $x^{\rm th}(t)$ represents the thermal-driven test mass motion. The frequency $\omega_{mc}$ is defined as $\omega_{mc}=\sqrt{\omega_m^2-\gamma_m^2/4}$.

In the steady state, the second-order conditional moments of the test mass converge to constants:
\begin{align}
    V_{xx}(+\infty) &=\frac{\hbar}{\sqrt{2}M\omega_q}\frac{1}{\sqrt{1+\sqrt{1+\Lambda_q^4}}},\label{Vxx_inf}\\
    V_{xp}(+\infty) &=\frac{\hbar}{2\Lambda_q^2}\left(-1+\sqrt{1+\Lambda_q^4}\right),\\
    V_{pp}(+\infty) &=\frac{\hbar M\omega_q}{\sqrt{2}}\frac{\sqrt{1+\Lambda_q^4}}{\sqrt{1+\sqrt{1+\Lambda_q^4}}},
\end{align}
where $\Lambda_q=\sqrt{\hbar\alpha^2/(M\omega_q^2)}$. Using Eq.~\eqref{eq:measurement result} and Eq.~\eqref{eq:evolution of x(classical noise)}, under the high quality factor $Q_m=\omega_m/\gamma_m \simeq \omega_{mc}/\gamma_m$ limit, the power spectrum density\,(PSD) of outgoing light can be derived as:

\begin{equation}
    \begin{split}
            S_{\tilde{y}\tilde{y}}^{(\rm{SN})}(\Omega)\simeq&\frac{1}{M^2((\Omega^2-\omega_m^2)^2+\gamma_m^2\Omega^2)}
            \Bigg[\hbar^2\alpha^4+4\alpha^2 M\gamma_m k_{b}T\\
            &-2M\omega_{\rm{SN}}^2\left(\sqrt{\hbar^2
            \alpha^4+M^2\omega_q^4}-M\omega_q^2\right)\Bigg]+1,
\end{split}
\end{equation}
where we have taken the high temperature limit\,($k_BT/\hbar\omega_m\gg1$). The radiation pressure of the light and thermal noise contribute to the first and second terms. The third term is the SN gravity effect in causal-condition prescription, and the last term is the shot noise. It is crucial to note that this spectrum shares the same shape as in standard QM\,(where $\omega_{\rm SN}=0$), differing only slightly in magnitude.

\begin{figure}[h]
    \includegraphics[scale=0.48]{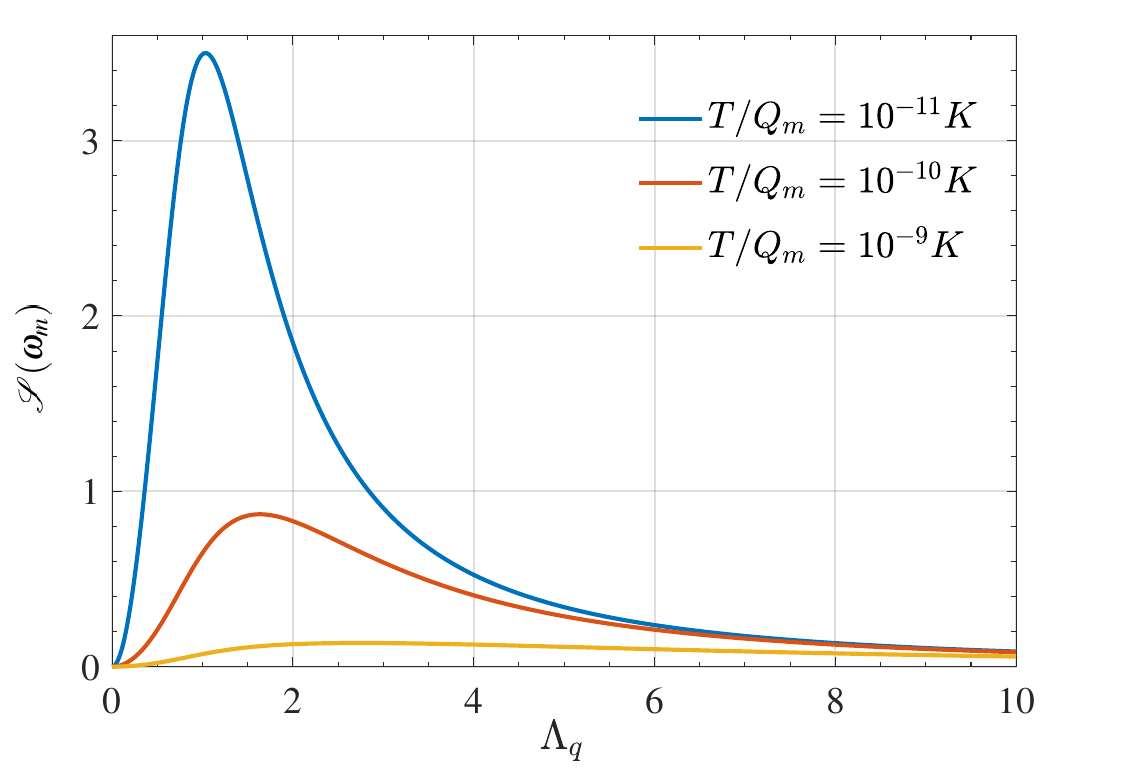}
    \caption{The logarithmic ratio $\mathcal{S}(\omega_m)$ of the spectrum of the outgoing light at the test mass’s eigenfrequency $\omega_m$ for different temperatures.}
    \label{fig:squeezing level}
\end{figure}

The following logarithmic ratio $\mathcal{S}(\omega)$ can be defined to quantify the difference between the SN spectrum and that of standard QM:
\begin{equation}\label{eq:logarithmic ratio}
    \begin{split}
            \mathcal{S}(\omega_m)&=-10\log_{10}\left[\frac{S_{\tilde{y}\tilde{y}}^{(\rm{SN})}(\omega_m)}{S_{\tilde{y}\tilde{y}}^{(\rm{QM})}(\omega_m)}\right]\\
           & \simeq-10\log_{10}\left[1-\frac{\omega_{\rm{SN}}^2}{\omega_q^2}\frac{2\Lambda_q^2}{(\sqrt{1+\Lambda_q^4}+1)(\Lambda_q^2+\lambda)}\right],
    \end{split}
\end{equation}
where $\lambda=4k_B T/(\hbar\omega_q Q_m)$. Figure~\ref{fig:squeezing level} illustrates the logarithmic ratio at various temperatures, calculated based on the parameters provided in Table\,\ref{tab:system_parameters}.  

\begin{table}[h]
    \centering
    \begin{tabular}{|c|c|c|}
    \hline
    \textbf{Parameters} & \textbf{Symbol} & \textbf{Value} \\\hline
 Cavity length & $L$ & $2$ m\\
    Mirror mass & $M$ & $0.2$ kg \\
    Mechanical damping & $\gamma_{m}$  & $4\times10^{-10}\times2\pi$ Hz \\
    Mirror eigenfrequency &$\omega_{m}$ & $4\times10^{-3}\times2\pi$ Hz\\
    SN frequency & $\omega_{\rm{SN}}$ & $8.19\times10^{-2}\times2\pi$  Hz\\
    Optical wavelength & $\lambda$ & $1064$ nm\\
    Finesse & $\mathcal{F}$ & $300$\\
        \hline
    \end{tabular}
    \caption{Key parameters of the optomechanical system with a single movable mirror under SN self-gravity.
}
    \label{tab:system_parameters}
\end{table}

The demanding requirement for distinguishing the SN predictions from QG predictions motivates us to explore an alternative optical spectrum in the CCSN optomechanical system, which can exhibit a distinct spectral shape. Previous work by Miki \emph{et al.}\,\cite{Miki2025} points out the use of delayed-measurement schemes, in which the collapse of the optomechanical joint quantum state does not promptly source the classical gravity.  In this work, by implementing the fact that the second-order conditional moments of the test mass evolve with different frequencies in SN theory\,($2\omega_q$) from QG\,($2\omega_m$) in the Riccarti equation, we propose to extract this difference from the non-stationary evolution before the second-order moments reaching steady values.


\section{The dynamic of second-order conditional moments}\label{section:Riccati equation}
First, we discuss the non-stationary dynamics of the second-order conditional moments by analyzing the Riccati equation.  By combining Eq.~\eqref{eq:Vxx_classical}, Eq.\eqref{eq:Vxp_classical} and Eq.\eqref{eq:Vpp_classical}, we can derive a third-order differential equation of $V_{xx}$ as:
\be
(\mathcal{L}^m_t+\mathcal{L}^\alpha_t)V_{xx}(t)=0,
\ee
where the free mechanical evolution part is
\begin{equation}
   \mathcal{L}^m_tV_{xx}(t)\equiv \frac{M^2}{2}\left(\dddot{V}_{xx}+4\omega_q^2\dot{V}_{xx}\right),
\end{equation}
and measurement-induced part is
\begin{equation}
    \begin{split}
  &  \mathcal{L}^\alpha_tV_{xx}(t)\equiv  \frac{\alpha^2}{2}\times\\
& \left[  M^2\left(4 \alpha^4 V_{xx}^4 + 7 \dot{V}_{xx}^2+ 4 V_{xx}^2 \left(  \omega_q^2 + 4 \alpha^2 \dot{V}_{xx} \right) + 6 V_{xx} \ddot{V}_{xx}\right) -\hbar^2\right].
    \end{split}
\end{equation}
Introducing the following dimensionless variables:
\begin{equation}
    h_1(t)=\frac{V_{xx}(t)}{V_{xx}^{\rm{vac}}},\ \ h_2(t)=\frac{2V_{xp}(t)}{\hbar},h_3(t)=\frac{V_{pp}(t)}{V_{pp}^{\rm{vac}}}
\end{equation}
where $V_{xx}^{\rm{vac}}=\hbar/(2M\omega_q)$  and $V_{pp}^{\rm{vac}}=\hbar M\omega_q/2$ are the position error and the momentum error of the ground state, respectively. We can transform the above terms to be 
\begin{equation}\label{eq:Vxx}
    \begin{split}
      &\mathcal{L}^m_th_1(t)\equiv \frac{\dddot{h}_1}{2}+2\omega_q^2\dot{h}_1,\\
     &\mathcal{L}^\alpha_th_1(t)\equiv  \Lambda_*^{2}\omega_q\times \\
     &\quad \left[\frac{\Lambda_*^{4}}{4\omega_q^4}h_1^4+h_1^2+\frac{7}{4\omega_q^2}\dot{h}_1^2+\frac{2\Lambda_*^{2}}{\omega_q^3}h_1^2\dot{h}_1+\frac{3}{2\omega_q^2}h_1\Ddot{h}_1-1\right],
    \end{split}
\end{equation}
where $\Lambda_*=\Lambda_q\omega_q$ characterises the measurement strength.

The structure of the nonlinear differential equation indicates that the free evolution component governs the oscillatory behavior of the conditional second-moments, with a frequency of $2\omega_q$. The measurement term, which consists of nonlinear contributions, drives the conditional second-order moments to converge to a constant over time. Now we will analyze the impact of the parameters $\Lambda_*$ and $\omega_q$ on the evolution of the second-order conditional moments. Additionally, the initial state of the test mass also plays a crucial role in shaping the dynamics of the conditional second-moments, which will also be discussed accordingly.

\begin{figure}[h]
    \includegraphics[scale=0.25]{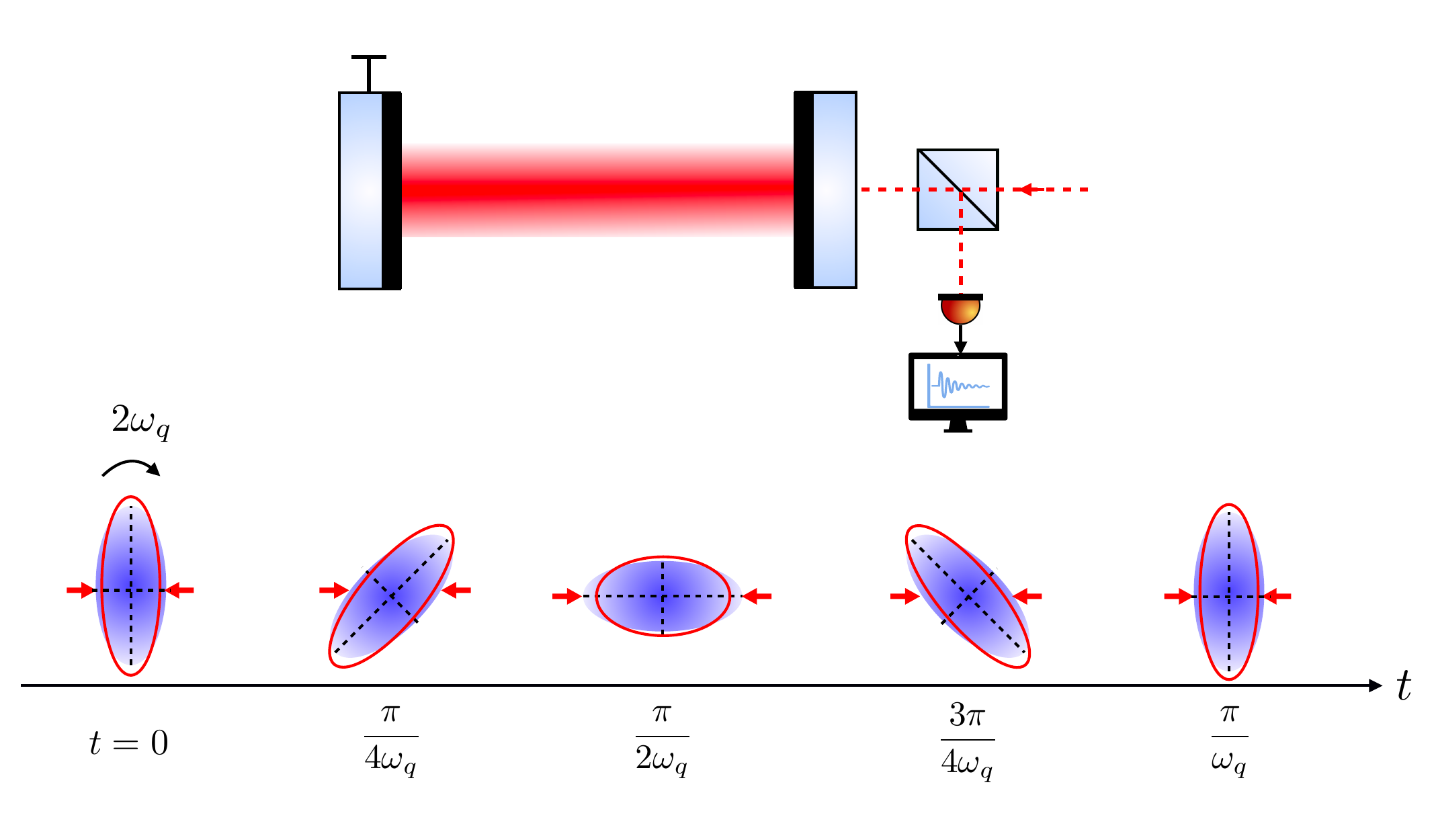}
    \caption{Schematic diagram of the system: Optomechanical interaction between the optical field and the movable end test mass mirror, where the test mass dynamics is influenced by the self-gravity of the test mass mirror sourced by the conditional mean displacement. The Wigner function of the test mass rotates with frequency $2\omega_q$, at the same time, the displacement measurement continuously prepares the conditional quantum state and drives the conditional mean displacement to the equilibrium value with time scale $\Lambda_*^2/\omega_q$. To have a distinct signature, the ratio $\Lambda_*^2/\omega^2_q$ should be smaller than one. }
    \label{fig:scheme}
\end{figure}

\subsection{The influence of measurement rate $\Lambda_*$ and oscillation frequency $\omega_q$}\label{subsec:the influence of measurement intensity}
To explore the influence of the measurement rate $\Lambda_*$ and the oscillation frequency $\omega_q$  on the evolution of the conditional variance, we first conduct a perturbative study around the equilibrium point. Setting $\dot{h_1}=\ddot{h_1}=\dddot{h_1}=0$, we can determine the equilibrium point of the equation as $h_1^{\rm{eq}} = V_{xx}(+\infty) / V_{xx}^{\rm{vac}}$. Around the equilibrium point, the equation can be linearized by letting $h_1 = h_1^{\rm{eq}} + \delta h_1$ and ignoring higher-order terms, such as $\delta h_1^2$, $\delta h_1 \dot{\delta h_1}$, etc, the resulting linearized equation is:
\begin{equation}\label{eq:linearized equation of evolultion of y1}
    \begin{split}
        \dddot{\delta h}_1+3\omega_q&\sqrt{2(\kappa-1)}\ddot{\delta h}_1+4\omega_q^2\left(2\kappa-1\right)\dot{\delta h}_1\\
        &+4\omega_q^3\kappa\sqrt{2(\kappa-1)}\delta h_1=0,
\end{split}
\end{equation}
where $\kappa=\sqrt{1+(\Lambda_*^2/\omega_q^2)^2}$. 

In the case of weak measurement $\Lambda_q^2\ll 1\,(\lesssim 0.13)$, this linearized equation has three roots,  approximately given as:
\be
\lambda_1\simeq-\frac{\Lambda_*^2}{\omega_q}, \quad \lambda_{2,3}\simeq-\frac{\Lambda_*^2}{\omega_q}\pm2i\omega_q.
\ee
The real parts of all three roots are negative, ensuring that the second-order moments converge to a stable value. The convergence rate is influenced by the measurement rate $\Lambda_*$ and the oscillation frequency $\omega_q$, increasing the measurement rate $\Lambda_*$ and decreasing the oscillation frequency $\omega_q$ both enhance the convergence of the conditional variance near the stable point.  The imaginary part of the roots exactly describes the oscillation frequency. In contrast, when the measurement strength is large,  the three roots become $\lambda_1\simeq-\sqrt{2}\Lambda_*$, $\lambda_2\simeq-\sqrt{2}(1+i)\Lambda_*$ and $\lambda_3\simeq-\sqrt{2}(1-i)\Lambda_*$, indicating that the measurement strongly modulates the oscillation frequency. Under this condition, it can be shown that no peak appears in the spectrum due to the strong decay rate. 

\begin{figure}[h]
    \includegraphics[scale=0.4]{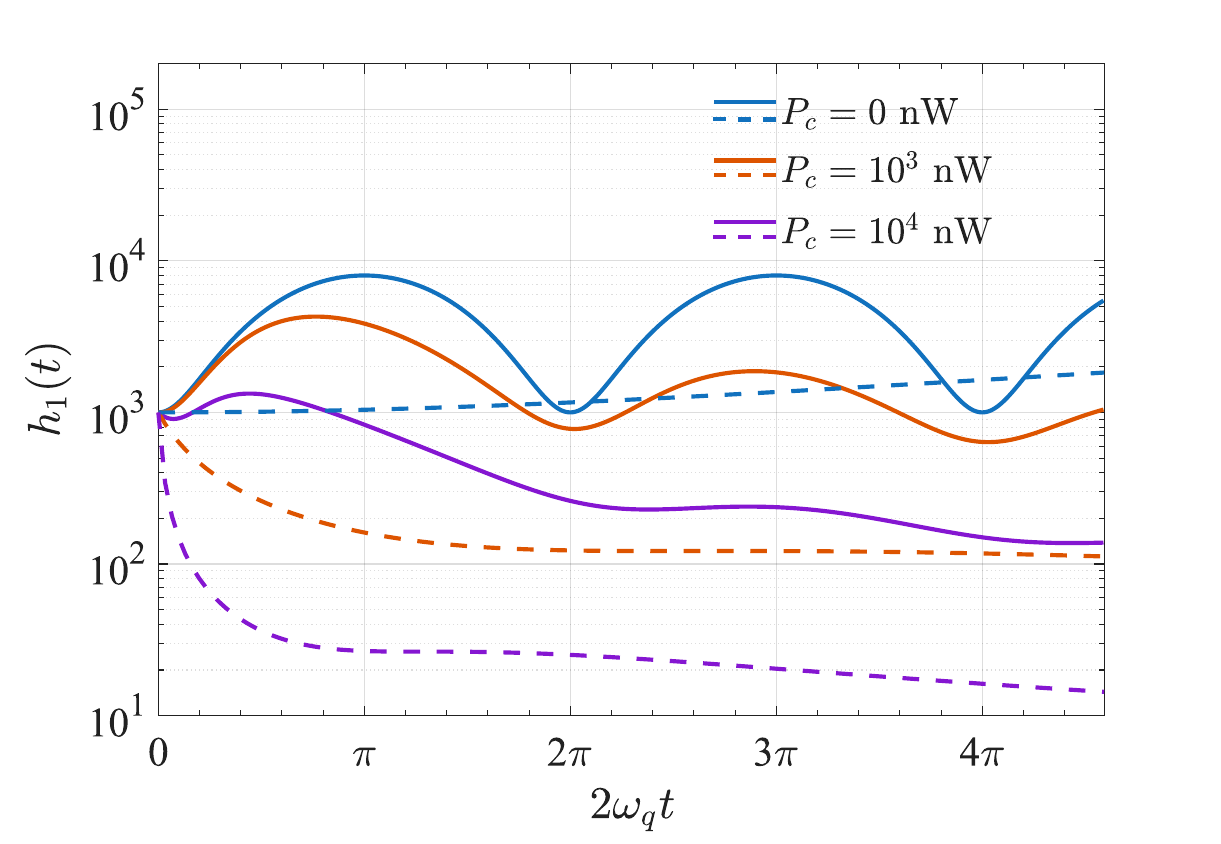}
    \includegraphics[scale=0.4]{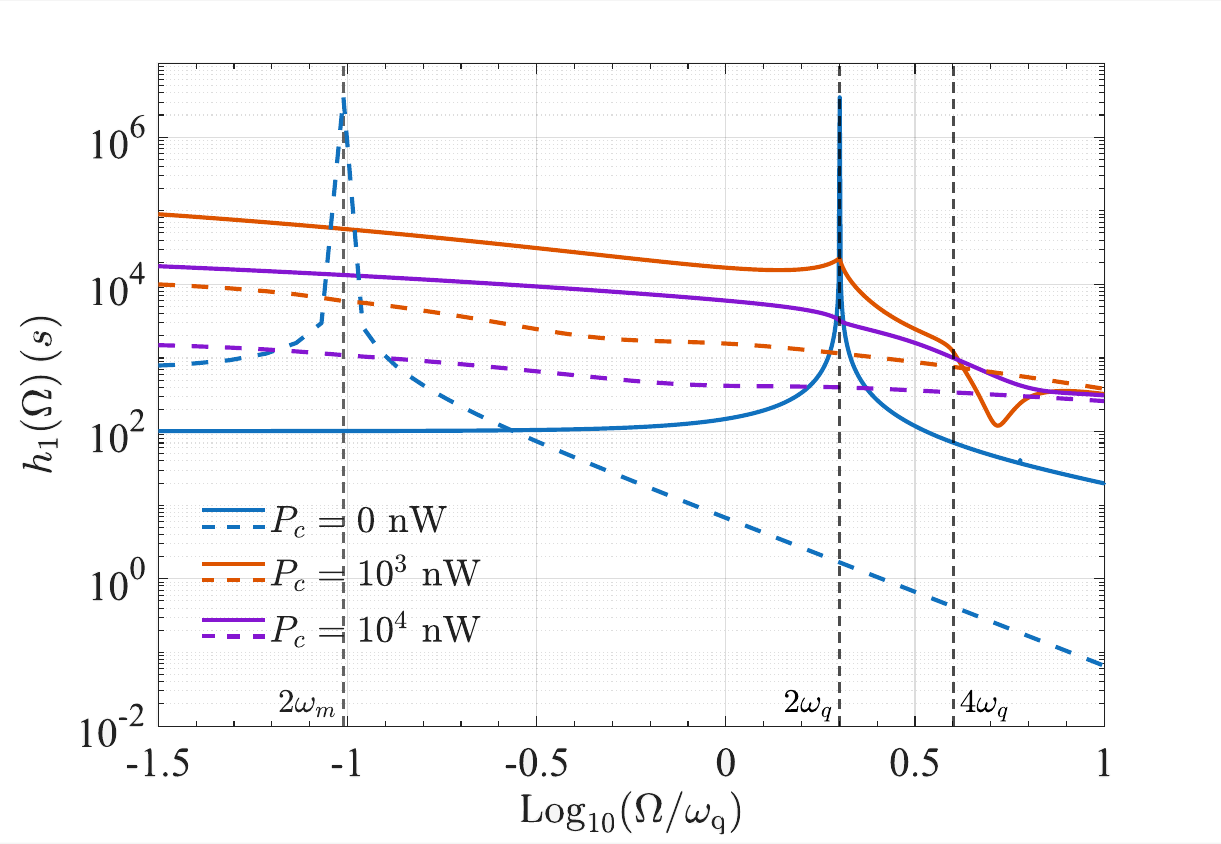}
    \caption{Time evolution of  $h_{1}$  under varying intra-cavity power and its corresponding frequency-domain behavior. The solid line is the SN case and the dash line correspond to the QG case.}
\label{fig:intensity influence of Vxx}
\end{figure}

\begin{figure}
   \includegraphics[scale=0.43]{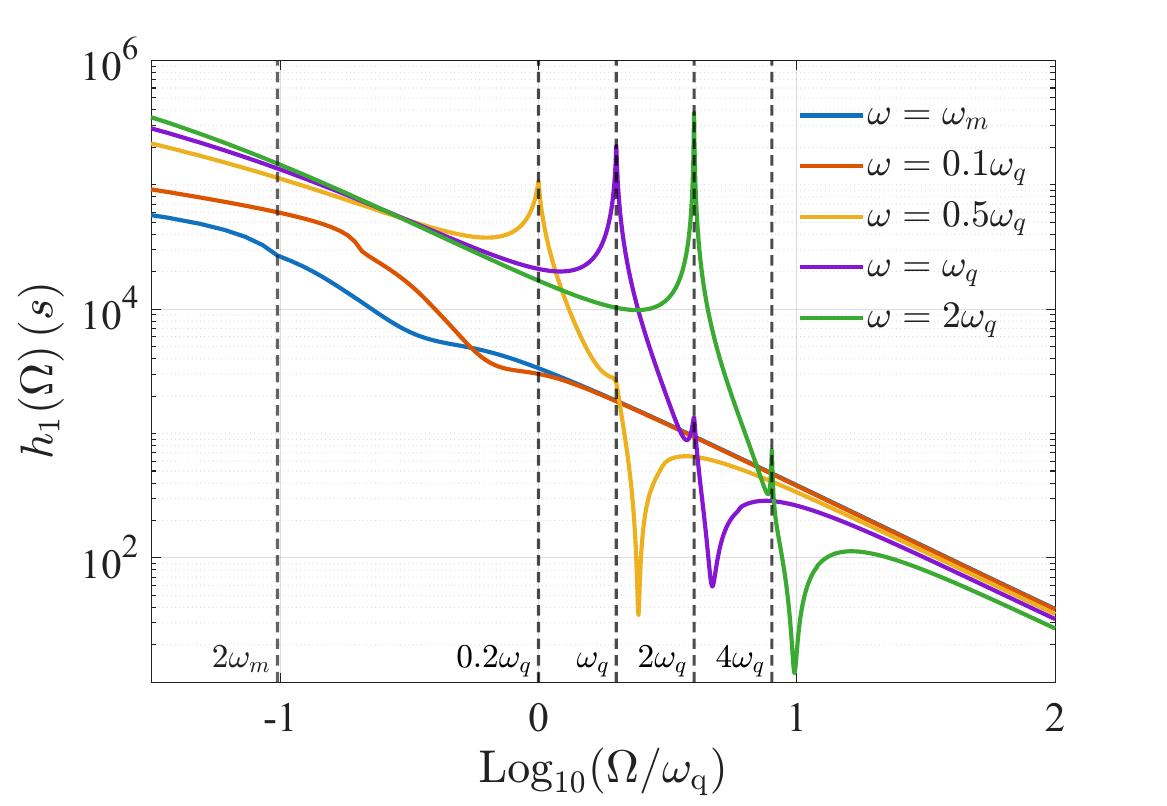}
\caption{Spectrum of $h_{1}$ under varying oscillation frequency.}
\label{fig:influence of frequency in freq domain}
\end{figure}

The above perturbative analysis shows that increasing the measurement rate $\Lambda_*$ and decreasing the oscillation frequency $\omega_q$ can enhance the conditional variance's convergence, suppressing the oscillatory behavior. To illustrate the effect of these parameters on the covariance dynamics beyond the perturbative region, we present the numerical results using the parameters listed in Table~\ref{tab:system_parameters} with an initial 4.5-dB squeezed mechanical state~\cite{santiagocondori2023} in\,Figure~\ref{fig:intensity influence of Vxx} and Figure~\ref{fig:influence of frequency in freq domain}: Figure~\ref{fig:intensity influence of Vxx} shows the influence of the intra-cavity power $P_{\rm cav}$ on $h_1(t)$ and its Fourier transform $h_1(\Omega)$; Figure~\ref{fig:influence of frequency in freq domain} illustrates the influence of the oscillation frequency $\omega_q$ on the spectrum of $h_1$, where the intra-cavity power is fixed to be $10^2$ nW. 

\begin{figure}[h]
    \includegraphics[scale=0.45]{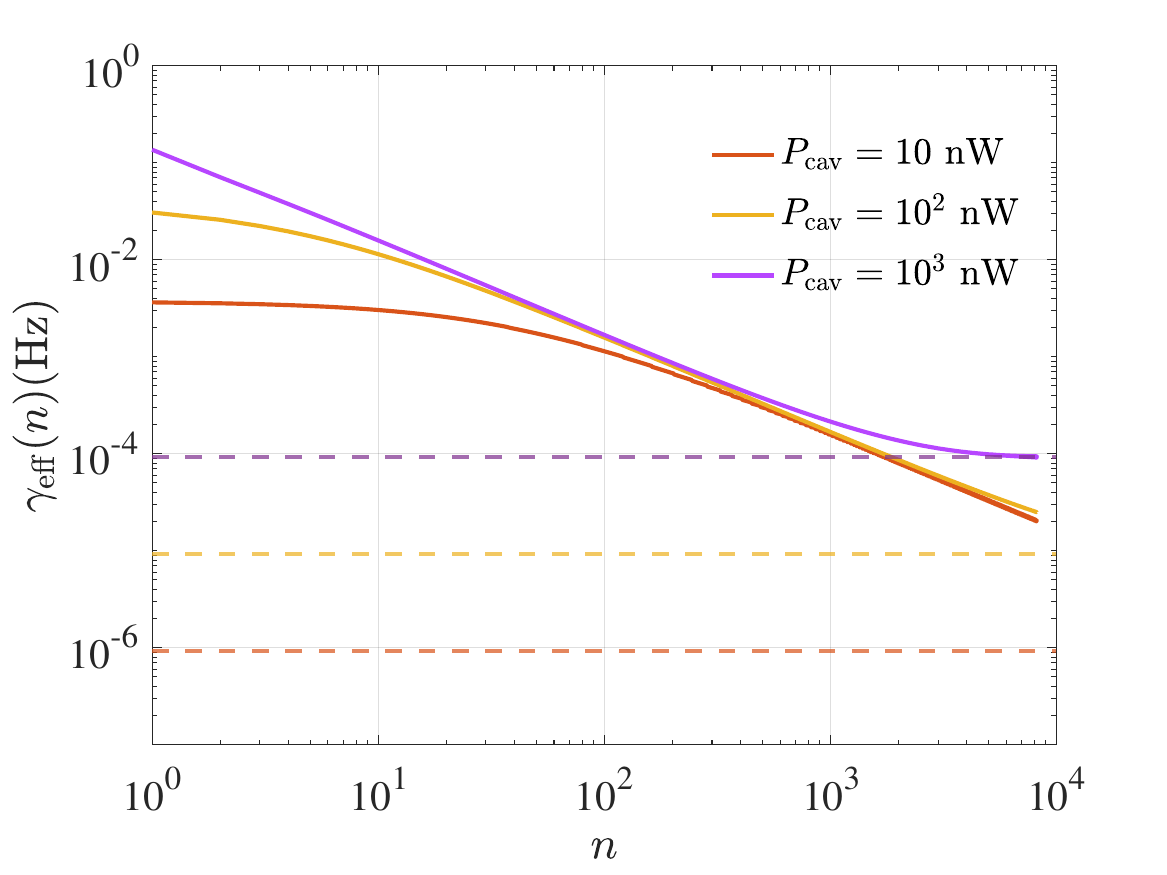}
    \caption{The effective decay rate in weak measurement situation. The solid line represents the equivalent decay rate for each period at different measured intensities. The dashed line represents the asymptotic decay rate near the equilibrium position.}
    \label{fig:effective decay rate}
\end{figure}

In these numerical results, besides the main peak at $2\omega_q$, the additional peak structure appearing at $4\omega_q$ is due to second-order corrections.  To characterize the effect of the non-perturbative terms on the  convergence rate, we define the following effective convergence rate assuming a weak measurement strength shown in Fig.~\ref{fig:effective decay rate}:
\begin{equation}
    \gamma_{\rm{eff}}(n)=\frac{\omega_q}{\pi}\log{\left[\frac{h_1(t_0+(n-1)\pi/\omega_q)-h_1^{\rm{eq}}}{h_1(t_0+n\pi/\omega_q)-h_1^{\rm{eq}}}\right]},
\end{equation}
where $t_0$ is time that $h_1(t)$ reaches its first peak, and $n$ is the peak's number.  The numerical results shown in Fig.\,\ref{fig:effective decay rate} indicate that at positions far from the equilibrium point, there is a higher convergence rate, while as the oscillation gradually approaches the equilibrium position, the convergence rate asymptotically goes to the theoretical result. 

The evolution of $V_{xp}$ and $V_{pp}$ can be similarly derived from the evolution of $V_{xx}$ by using Riccati equation, the influence of the measurement rate $\Lambda_*$ and the oscillation frequency $\omega_q$ on the evolution of $V_{xp}$ and $V_{pp}$ is similar to that of $V_{xx}$, which will not be repeated here.

\subsection{The choice of the initial state}

In our system, the conditional state maintains its Gaussianity throughout the evolution process, of which its uncertainty distributions are represented by an error ellipse in phase space, i.e. a cross-section of the Wigner function $\mathcal{W}(x,p)$. This ellipse can be fully characterized by three parameters: the squeezing degree $r$, the squeezing angle $\theta$, and the scale factor $\beta$. After normalizing the position and momentum operators, the equation of the uncertainty ellipse in phase space can be expressed as:
\begin{equation}
    \begin{split}
        (\cosh2r+\cos2\theta\sinh2r)\hat{X}^2&+(\cosh2r-\cos2\theta\sinh2r)\hat{P}^2\\
        &+\sinh2r\sin2\theta\hat{X}\hat{P}=\beta
    \end{split}
\end{equation}
where $\hat{X}=\hat{x}/\sqrt{V_{xx}^{\rm{vac}}}$ and $\hat{P}=\hat{p}/\sqrt{V_{pp}^{\rm{vac}}}$ are the normalized position and momentum operators. The relationships between our dimensionless variables $h_1$, $h_2$, $h_3$ and these ellipse parameters can be established as follows.

The projections of the ellipse onto the $X$ and $P$ axes respectively correspond to $h_1$ and $h_3$ as:
\begin{align}
    h_1&=\beta\left[\cosh 2r-\cos2\theta\sinh2r\right],\\
    h_3&=\beta\left[\cosh2r+\cos2\theta\sinh2r\right].
\end{align}
Meanwhile, $h_2$ relates to the correlation between position and momentum, or equivalently, to the projections along the $\pi/4-$rotated axi\,(or the diagonal line):
\begin{equation}
    h_2=\frac{\langle\hat{X}\hat{P}+\hat{P}\hat{X}\rangle_c}{2}=\frac{\langle\hat{X'}^2-\hat{P'}^2\rangle_c}{2}=\beta\sin2\theta\sinh{2r},
\end{equation}
where $\hat{X'}=(\hat X+\hat P)/\sqrt{2}$ and $\hat{P'}=(-\hat X+\hat P)/\sqrt{2}$ represent operators rotated by $\pi/4$ in phase space.

Considering the time evolution, we further parameterize these quantities by setting $r\rightarrow r(t)$, $\theta\rightarrow\theta(t)$ and $\beta\rightarrow\beta(t)$:
\begin{align}
    h_1(t)&=\beta(t)\left[\cosh 2r(t)-\cos2\theta(t)\sinh2r(t)\right],\label{eq:y1 with parameterized}\\
    h_2(t)&=\beta(t)\sin2\theta(t)\sinh{2r(t)},\label{eq:y2 with parameterized}\\
    h_3(t)&=\beta(t)\left[\cosh2r(t)+\cos2\theta(t)\sinh2r(t)\right].\label{eq:y3 with parameterized}
\end{align}
By substituting Eqs.\eqref{eq:y1 with parameterized}--\eqref{eq:y3 with parameterized} into the Riccati equation, we obtain the following dynamical equations for the parameters $r(t)$, $\theta(t)$ and $\beta(t)$:
\begin{align}
    \dot{r}&=\frac{\Lambda_*^2}{2\omega_q}(\cos2\theta\cosh2r-\sinh2r)\frac{1+\beta^2}{2\beta},\label{eq:evolultion of r}\\
    \dot{\theta}&=\omega_q-\frac{\Lambda_*^2}{2\omega_q}\csch2r\sin2\theta\frac{1+\beta^2}{2\beta},\label{eq:evolultion of theta}\\
    \dot{\beta}&=\frac{\Lambda_*^2}{2\omega_q}\left(\cosh2r-\cos2\theta\sinh2r\right)(1-\beta^2).\label{eq:evolultion of mu}
\end{align}

Next, we will analyze the evolution of $h_1(t)$ for different ranges of the initial parameters $r(0),\beta(0)$.

\subsubsection*{Case 1: a Gaussian state with $r(0)\simeq0$}
When the $r(0)\simeq0$, the cross section of the initial state's the Wigner function $\mathcal{W}(x,p)$ is a circle. Since the measurement strength is weak, the cross-section of the final stationary state's Wigner function will be only slightly squeezed. Therefore, during the non-stationary evolution process, the signature amplitude is weak, hence, an initial Gaussian state with relatively strong squeezing is necessary.

\subsubsection*{Case 2: a squeezed  Gaussian state  with $r(0)>0$}
In this case, the ratio between the two terms on the right-hand-side of Eq.\eqref{eq:evolultion of theta} satisfies:
\begin{equation}\label{eq:rotation condition}
    \frac{\Lambda_*^2}{2\omega^2_q}\csch2r\sin2\theta\frac{1+\beta^2}{2\beta}\sim   \frac{\Lambda_*^2}{\omega^2_q}e^{-2r}\frac{1+\beta^2}{2\beta},
    \end{equation}
where we should note that previously we have shown that $\Lambda_*/\omega_q$ should be small to keep the oscillatory behavior of the conditional second-order moments. With $r>0$ and setting $\beta(0)\gg1$ but keeping the above ratio much smaller than one,  we then have the approximated solution of $\theta(t)$ as $\theta(t)=\omega_q t+\theta_0$ and the evolution of initial state parameters Eqs.\,\eqref{eq:evolultion of r}-\eqref{eq:evolultion of mu} can be approximatted as:
\begin{equation}
    \begin{split}
        \dot{r}&=\frac{\Lambda_*^2}{4\omega_q}e^{2r}[\cos(2\theta_0+2\omega_qt)-1]\frac{\beta}{2},\\
        \dot{\beta}&=-\frac{\Lambda_*^2}{4\omega_q}e^{2r}[\cos(2\theta_0+2\omega_qt)-1]\beta^2.
    \end{split}
\end{equation}
The above equations can be analytically solved as:
\begin{equation}
    \begin{split}
        r(t)&=r(0)-\frac{1}{4}\log\left[1+e^{2r(0)}\beta(0)g(t)\right],\\
        \beta(t)&=\beta(0)/\sqrt{1+e^{2r(0)}\beta(0)g(t)},
    \end{split}
\end{equation}
with
\begin{equation}
    \begin{split}
        g(t)\equiv\frac{\Lambda_*^2}{2\omega_q^2}[\omega_qt-\cos(\omega_qt+2\theta_0)\sin\omega_qt]\ge0.
    \end{split}
\end{equation}
Finally, the evolution of $h_1(t)$ is,
\begin{equation}\label{eq:y1 with mu large}
    h_1(t)=\frac{\beta(0)^2\cos^2(\theta_0+\omega_qt)g(t)+h _1^{(0)}(t)}{1+e^{2r(0)}\beta(0)g(t)},
\end{equation}
where $h^{(0)}_1(t)$ is the evolution of $h_1$ without measurement:
\begin{equation}
    h^{(0)}_1(t)=\beta(0)\left[\cosh 2r(0)-\cos(2\theta_0+2\omega_qt)\sinh2r(0)\right].
\end{equation}
It is clear that $h_1(t)$-evolution exhibits a decaying oscillatory behavior.  Let the oscillation term $\cos(2\theta(0)+2\omega_qt)\rightarrow-1$, we can obtain the the upper envelope of $h_1(t)$ as
\begin{equation}
        h^{\rm max}_1(t)=\left[\frac{\Lambda_*^2}{2\omega_q}t+\frac{e^{-2r(0)}}{\beta(0)}\right]^{-1},
\end{equation}
which demonstrates that a small ratio $e^{-2r(0)}/\beta(0)$ leads to a larger signal amplitude.
However, the decay rate of the envelope $dh^{\rm max}_1(t)/dt$ is:
\be
\frac{dh^{\rm max}_1(t)}{dt}=-\frac{\Lambda_*^2}{2\omega_q}\left[\frac{\Lambda_*^2}{2\omega_q}t+\frac{e^{-2r(0)}}{\beta(0)}\right]^{-2},
\ee
which takes its maximum value around the initial time $t=0$.  The faster the decay rate, the less signature in its WV spectrum\,(see the next section), hence a larger $e^{-2r(0)}/\beta(0)$ is favored. Therefore, the parameters $\beta(0), r(0)$ should be chosen in the way so that both the large signal amplitude and the small decay rate are considered.  Practically preparing a highly-squeezed vacuum mechanical state is very difficult; therefore, a squeezed thermal state such as the one prepared in the work by Santiago-Condori et al.\,\cite{santiagocondori2023} can be a potential choice. Usually, for a squeezed vacuum, the squeezing decibel scale is defined by the ratio between the semi-minor-axis value of the Wigner function's cross-section and that of the ground state value. However, for a squeezed thermal state, we define a new decibel scale to quantify the squeezing thermal state as:
\begin{equation}
  \text{Sqz dB}=  -\frac{1}{2}\times 10\log_{10}\left[\frac{h_{1}+h_{3}-\sqrt{(h_1-h_3)^2+4h_2^2}}{h_{1}+h_{3}+\sqrt{(h_1-h_3)^2+4h_2^2}}\right],
\end{equation}
where the ratio in the square bracket is the ratio between the the semi-major-axis value  and the semi-minor-axis value of the Wigner function's cross-section. Using this definition, the state prepared in\,\cite{santiagocondori2023} has $\beta(0)=2.8\times10^3$ and the squeezing level of 4.5\,dB.
\begin{figure}[h]
   \includegraphics[scale=0.43]{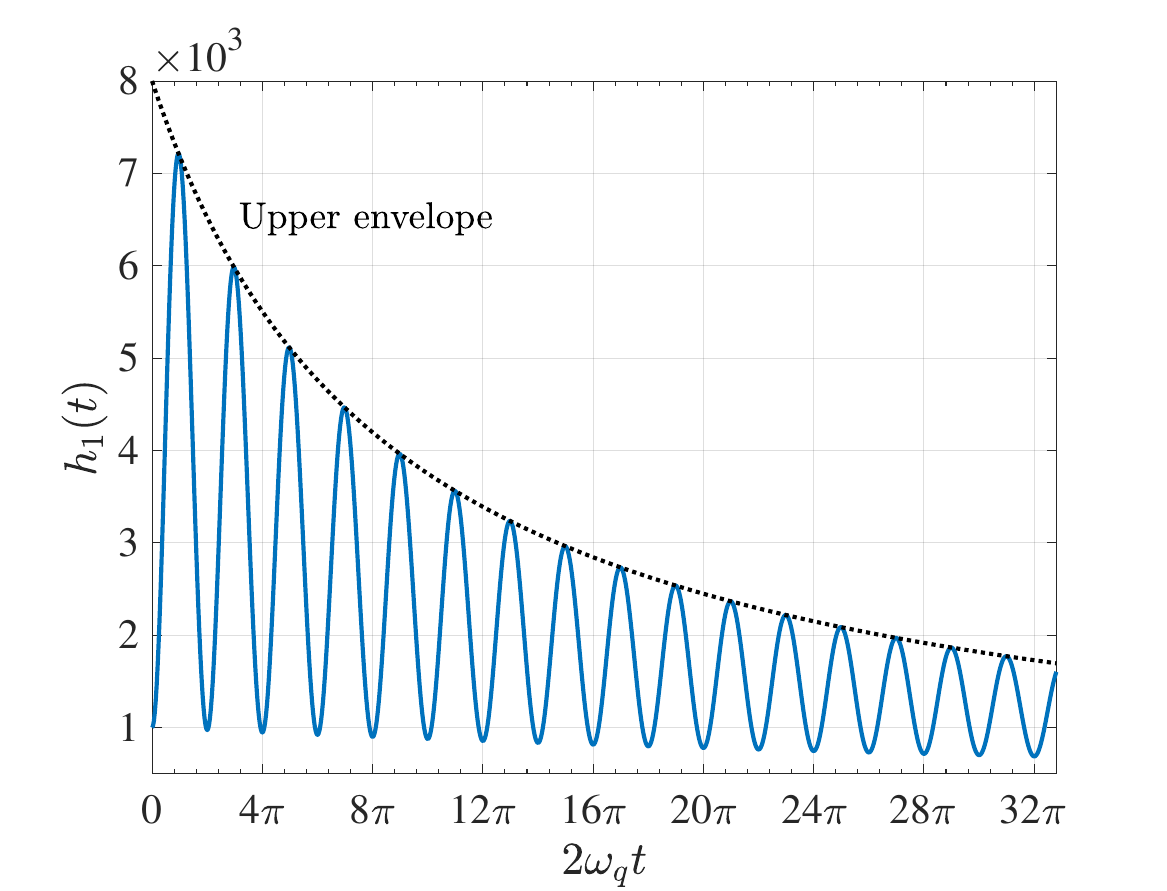}
   \caption{Time evolution of $h_1$ with the 4.5\,dB squeezed thermal state. The dotted line is the upper envelope calculated theoretically.} 
\label{fig:the influence of squeezing level}
\end{figure}

\begin{figure}[h]
    \includegraphics[scale=0.41]{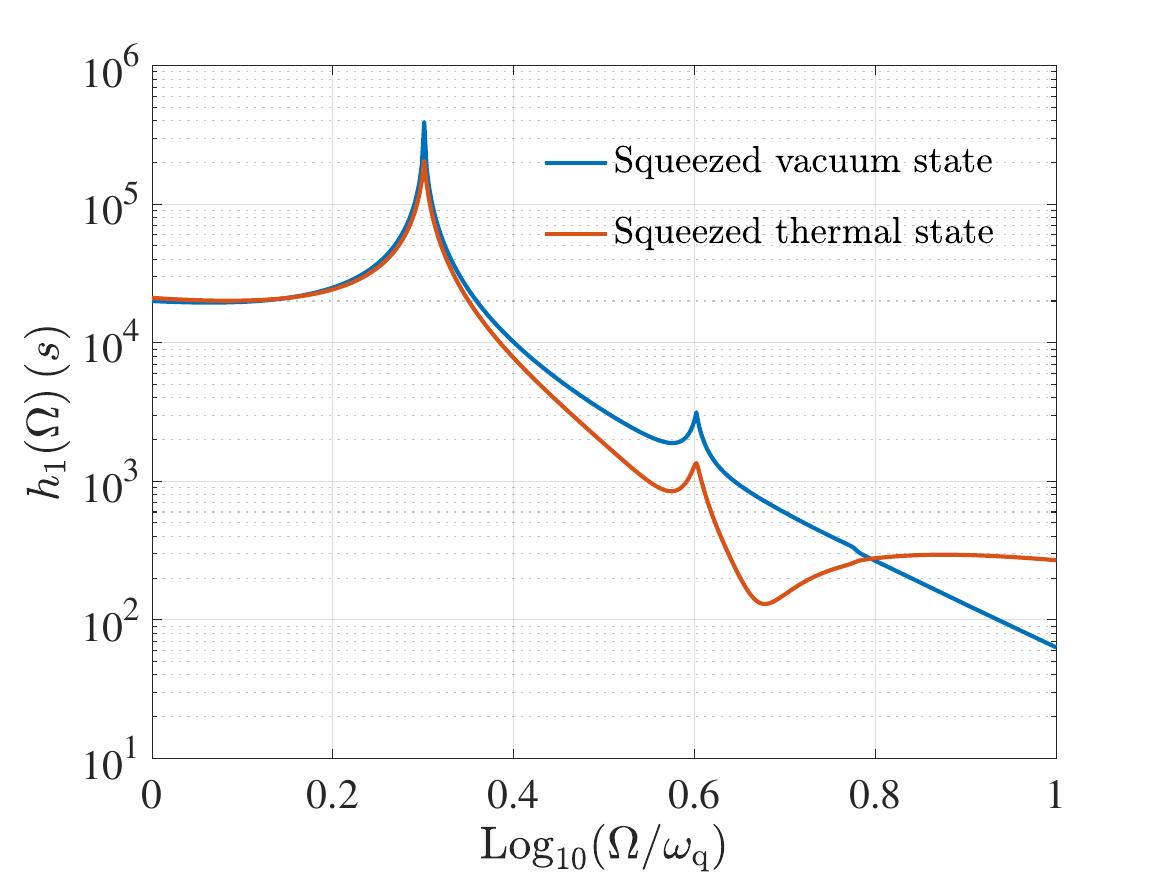}
    \caption{Comparison of the optimal squeezing vaccum state and squeezed thermal state when the intra-cavity power is $100$\,nW.}
\label{fig:CVS_vs_CSS}
\end{figure}

The numerical results shown in Figure~\ref{fig:CVS_vs_CSS} confirms that the 4.5\,dB squeezed thermal state as prepared in\,\cite{santiagocondori2023} can have comparable effect on the spectrum when a 39\,dB squeezed vacuum is used, which is currently impossible to prepare. In Figure~\ref{fig:CVS_vs_CSS}, the peak value in the spectrum of $h_1$ for the 4.5\,dB squeezed thermal state reaches approximately $0.52$ times that of the 39\,dB squeezed vacuum. In the subsequent calculations, we will take this squeezed thermal state as the benchmark.

\section{The signature of self-gravity in non-stationary process}\label{section:non-stational spectrum}
The above-discussed evolution of the second-order conditional moments in the SN theory exhibits distinctive features, which can be extracted by analyzing the optical data. This section develops a framework for analyzing the SN features from the non-stationary optomechanical dynamics of the test mass.

For a stationary Gaussian stochastic process $\tilde{x}(t)$ observed over a finite time interval $T_{\rm obs}$, its feature can be extracted by the power spectral density\,(PSD) defined as,
\begin{equation}
    \begin{split}
        S_{\tilde{x}\tilde{x}}(\Omega)&\equiv\frac{1}{T_{\rm obs}}\mathbb{E}[|\tilde{x}(\Omega)|^2],
    \end{split}
\end{equation}
where $\tilde{x}(\Omega)$ represents the Fourier transform of $\tilde{x}(t)$, and $\mathbb{E}[\cdot]$ denotes the ensemble average over all possible realizations. The physical interpretation of PSD is the distribution of noise power among different frequencies. 

The collected experimental data of a Gaussian stochastic process is a time series $\tilde{x}(t)$, with the mean value as the first-order moments and the correlation function as the second-order moment:
\begin{equation}
    C_{\tilde{x}\tilde{x}}(t,\tau)=\mathbb{E}\left[\tilde{x}(t+\tau)\tilde{x}^*(t)\right].
\end{equation}
For a stationary stochastic process, the correlation function depends only on the time interval $\tau$, i.e., $C_{\tilde{x}\tilde{x}}(t,\tau)=C_{\tilde{x}\tilde{x}}(\tau)$, which is related to the PSD via the Wiener-Khinchin theorem: 
\begin{equation}
    S_{\tilde{x}\tilde{x}}(\Omega)=\int_{-T_{\rm obs}}^{T_{\rm obs}}d\tau\ C_{\tilde{x}\tilde{x}}(\tau)e^{-i\Omega\tau}.
\end{equation}

In contrast, for non-stationary processes, $C_{\tilde{x}\tilde{x}}(t,\tau)\neq C_{\tilde{x}\tilde{x}}(\tau)$, as the correlation depends both on the time interval $\tau$ and the absolute time $t$. Consequently, the Wiener-Khinchin theorem no longer applies, and the conventional PSD fails to capture the feature of the second-order moments in our system, which can be shown by combining the Fourier transformation of Eq.~\eqref{eq:the evolultion of the x(classical noise)} and Eq.~\eqref{eq:the evolultion of the p(classical noise)} and obtain,
\begin{equation}
    S_{xx}(\Omega)= \frac{2\alpha^2 |G_m(\Omega)|^2}{T_{\rm obs}}\int_{0}^{T_{\rm obs}} dt\ |M(\gamma_m-i \Omega)V_{xx}(t)+V_{xp}(t)|^2,
\end{equation}
where $G_m(\Omega)=1/[M(\omega_m^2-\Omega^2-i\gamma_m\Omega)]$ is the mechanical response function. This integral effectively averages out the oscillatory behavior of the conditional variance, resulting in a spectrum with only a single peak at frequency $\omega_m$. The same conclusion applies to the spectrum of the output optical data $\tilde{y}(t)$. Therefore, to properly extract the oscillatory behavior of the conditional variance in the non-stationary SN evolution, an alternative spectral analysis framework specifically designed for non-stationary processes is needed.

\subsection{Wigner-Ville spectrum}
\begin{figure*}
        \includegraphics[scale=0.4]{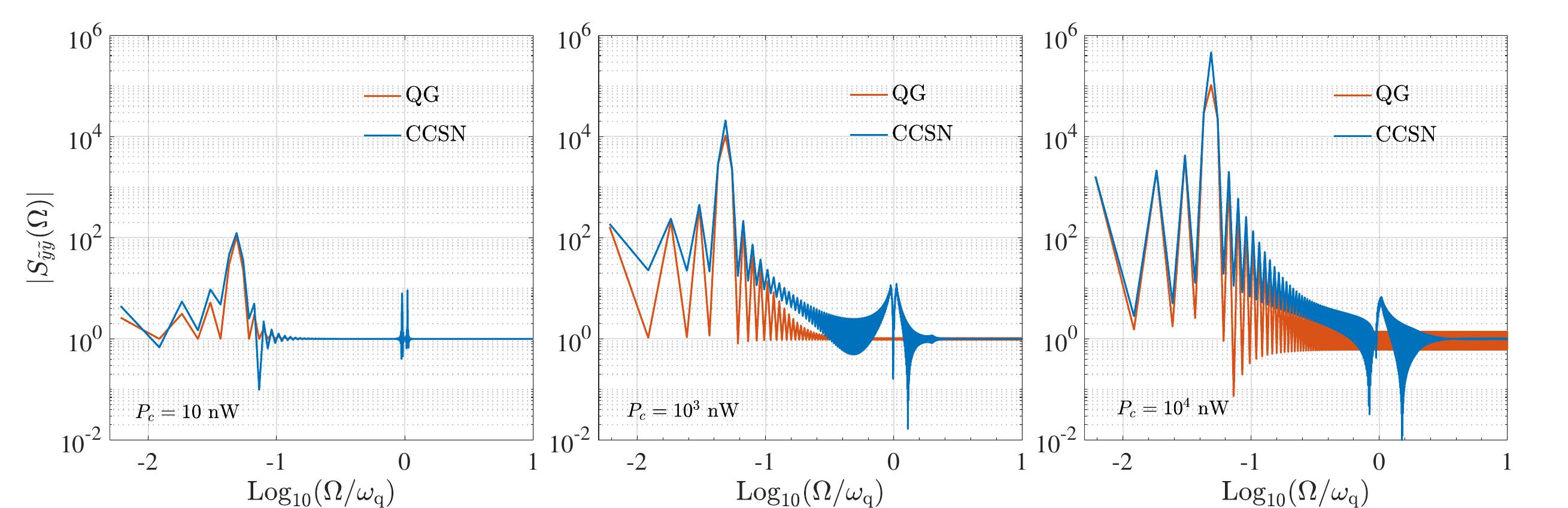}
        \caption{The WV spectrum of the outgoing light quadrature $\tilde{y}$ in $10^3$ sec experimental time under different cavity powers. At optimal power\,($P_{cav}=10^2\sim10^3$ nW), distinctive peaks at frequencies $\omega_q\pm\omega_m$ emerge in SN gravity\,(blue curve) that are absent in standard quantum mechanics\,(red curve), providing a clear experimental signature to distinguish between the two theories. }
        \label{fig:spectrum of yy}
\end{figure*}

The Wigner-Ville\,(WV) spectrum—an extension of the Wigner distribution developed by Ville\,\cite{PhysRev.40.749,Ville1948,10.5555/200604}—provides a powerful framework for the non-stationary process. Here we will show that WV spectrum can help reveal distinctive peaks at frequencies $\omega_q\pm\omega_m$ in CCSN theory that are notably absent in QG, which offer a clear experimental signature for distinguishing between the two theories. 

Suppose we have an experiment that is repeated for $N$ times, an ensemble of stochastic real signal $y_i(t)$\,($i\in[1,N]$) is generated. By averaging over all the realizations and taking the limit as $N\rightarrow\infty$, we can construct a correlation function of the signal $\tilde{y}(t)$ as:
\begin{equation}\label{eq:correlation function defination}
\begin{split}
    C_{\tilde{y}\tilde{y}}(t,\tau)&\equiv\lim_{N\rightarrow\infty}\frac{1}{N}\sum_{i=1}^N y_i\left(t+\frac{\tau}{2}\right)y_i\left(t-\frac{\tau}{2}\right)\\
   & =\mathbb{E}\left[\tilde{y}\left(t+\frac{\tau}{2}\right)\tilde{y}\left(t-\frac{\tau}{2}\right)\right].
\end{split}
\end{equation}\\
If the stochastic signal is stationary, it reduces to the conventional correlation function $C_{\tilde{y}\tilde{y}}(\tau)$. Moreover, it preserves the even symmetry with respect to $\tau$, a property that will be useful in defining a generalized spectrum. 

Next, we extend the Wiener-Khinchin theorem to non-stationary signals by defining a new spectrum. Similar to the stationary case, we perform a Fourier transformation of the above correlation function for $\tau$, defining what is known as the WV spectrum:
\begin{equation}\label{eq:WV spectrum defination}
    S^{\rm{WV}}_{\tilde{y} \tilde{y}}(t,\Omega)\equiv\int_{-\infty}^{\infty}\frac{d\tau}{2\pi}\ C_{\tilde{y}\tilde{y}}(t,\tau)e^{-i\Omega\tau}.
\end{equation}
It is important to note that the WV spectrum does not necessarily yield non-negative values at all frequencies—a fundamental limitation connected to the uncertainty principle, which prohibits simultaneously precise measurements of both time and frequency information. Consequently, the WV spectrum should be interpreted as a quasi-power spectrum, analogous to quasi-probability distributions\,(e.g. Wigner function) in quantum mechanics.

Substituting Eq.~\eqref{eq:evolution of x(classical noise)} and Eq.~\eqref{eq:measurement result} into Eq.~\eqref{eq:correlation function defination}, the correlation function of the optical phase measurement data in SN gravity during the non-stationary evolution is:
\begin{equation}
\begin{split}
C_{\tilde{y}\tilde{y}}(t,\tau)=&\alpha^2 \left[C_{xx}^{m}(t,\tau)+C_{xx}^{\rm th}(t,\tau)\right]+ \alpha C_{xdW}(t,\tau)\\
    &+C_{dW}(t,\tau),
\end{split}
\end{equation}
in which we have:
\begin{widetext}
    \begin{equation}\label{eq:correlation of xx}
    \begin{split}
        &\alpha^2C^{m}_{xx}(t,\tau)=\frac{\Lambda_*^4}{2\omega_q^2}\int_0^{t-\frac{|\tau|}{2}}ds\ e^{\gamma_m(s-t)}\left[\frac{\omega_m}{\omega_{mc}}h_1(s)\cos\omega_{mc}\left(s-t-\frac{\tau}{2}-\phi\right)-\frac{\omega_q}{\omega_{mc}}h_2(s)\sin\omega_{mc}\left(s-t-\frac{\tau}{2}\right)\right]\times\left[\tau\rightarrow-\tau\right], \\
        &\alpha^2C^{\rm th}_{xx}(t,\tau)=\Lambda_*^2\frac{ \omega_m^2}{2\omega_{mc}^2Q_m}\coth\left( \frac{\hbar \omega_m}{2 k_B T} \right)\left[e^{-\gamma_m\frac{|\tau|}{2}}\left(\frac{\cos(\omega_{mc}\tau)}{\gamma_m}+\frac{\sin(\omega_{mc}|\tau|+\phi)}{2\omega_m}\right)-e^{-\gamma_m t}\left(\frac{\cos(\omega_{mc}\tau)}{\gamma_m}+\frac{\sin(2\omega_{mc}t+\phi)}{2 \omega_{m}}\right)\right],\\
        &\alpha C_{xdW}(t,\tau)=\frac{\Lambda_*^2}{2\omega_q} e^{-\gamma_m\frac{|\tau|}{2}}\left[\frac{\omega_m}{\omega_{mc}}h_1\left(t-\frac{|\tau|}{2}\right)\cos(\omega_{mc}|\tau|+\phi)+\frac{\omega_q}{\omega_{mc}}h_2\left(t-\frac{|\tau|}{2}\right)\sin(\omega_{mc}|\tau|)\right]{\rm sgn}(2t-|\tau|),\\
        &\ C_{dW}(\tau)=\frac{1}{2}\delta(\tau),
\end{split}
\end{equation}
\end{widetext}
where ${\rm sgn}(x)$ is the sign function and the loss angle is defined as $\tan\phi=-\gamma_m/(2\omega_{mc})$. In obtaining the above result, we have used the correlation of the Wiener increment 
\be
\mathbb{E}\left[dW(t)dW(t+\tau)\right]=dt\delta(\tau).
\ee

\begin{figure*}
    \includegraphics[scale=0.4]{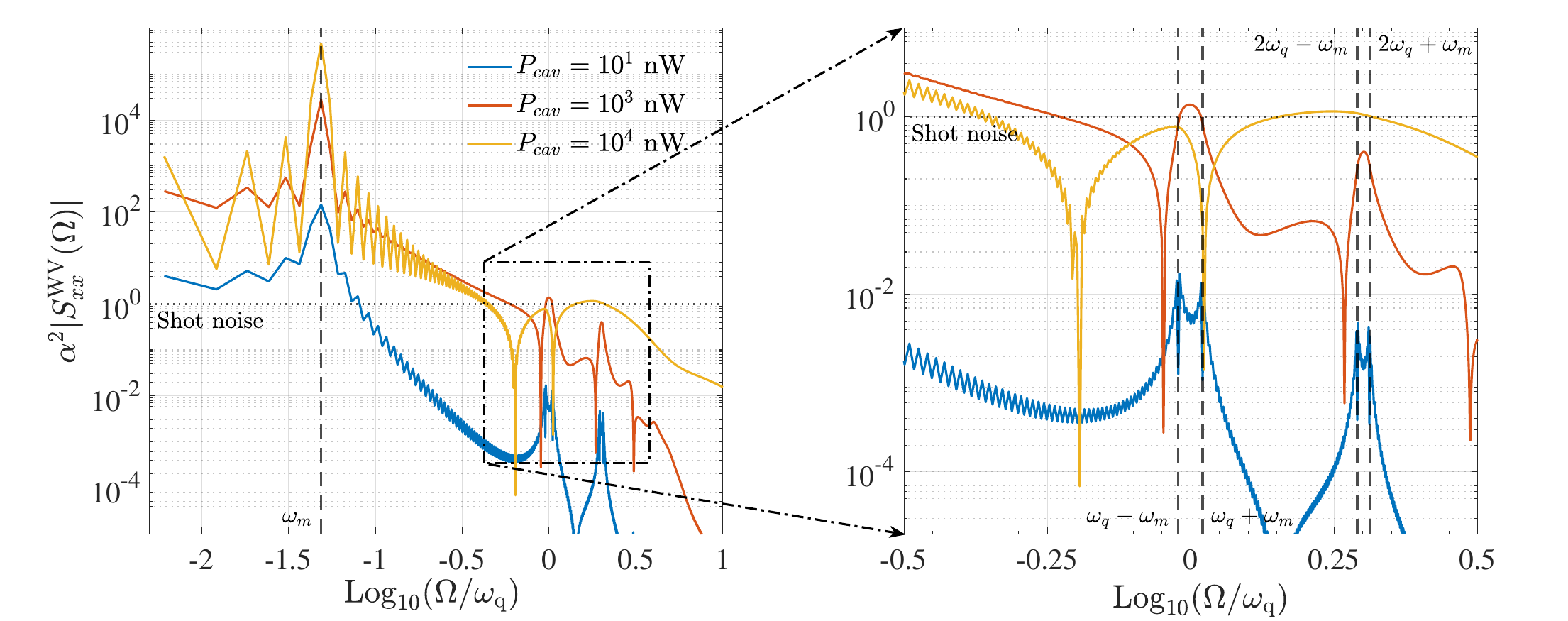}
    \includegraphics[scale=0.4]{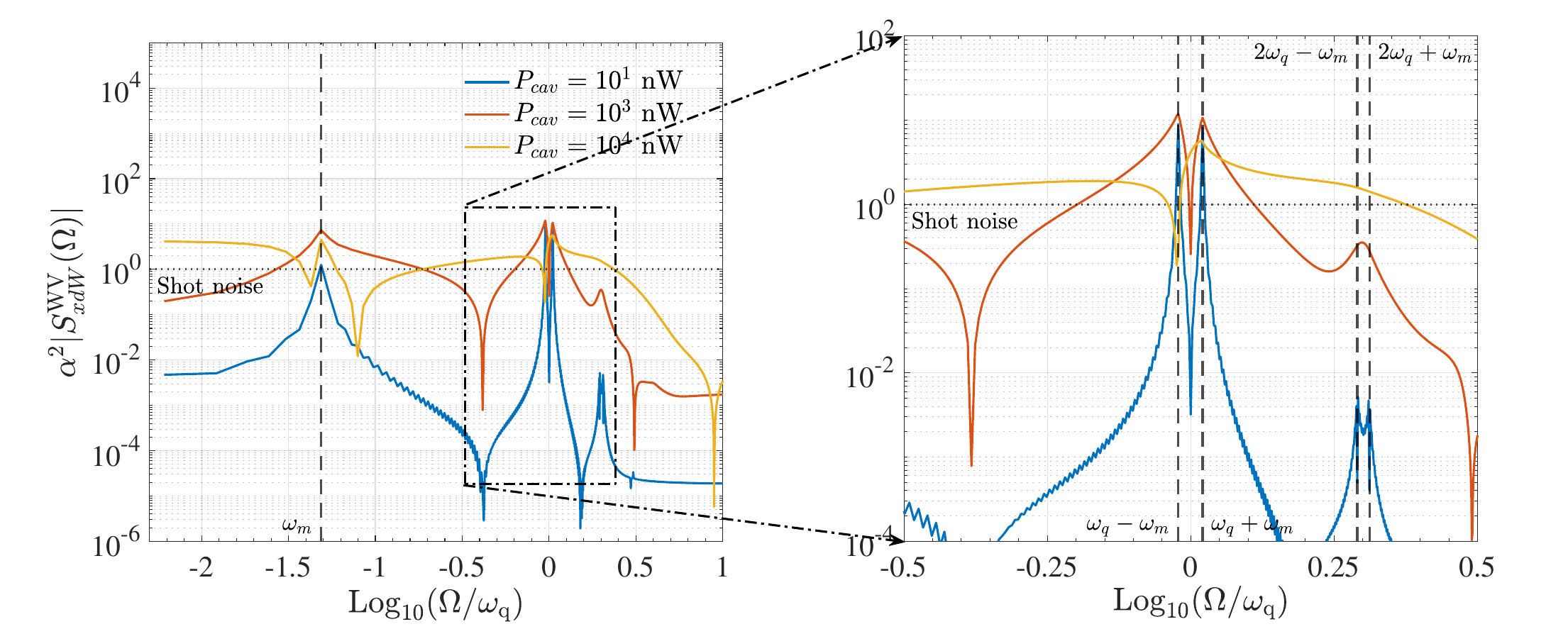}
    \caption{WV spectrum analysis of correlation components with $T/Q_m=10^{-10}$ K. Upper panel: The WV spectrum of position autocorrelation $C_{xx}(t,\tau)$, revealing characteristic peaks at frequencies $\omega_q \pm \omega_m$ that are unique to SN gravity. Lower panel: The WV spectrum of position-measurement noise cross-correlation $C_{xdW}(t,\tau)$, exhibiting similar spectral features.}
    \label{fig:spectrum of xx and xdW}
\end{figure*}

Finally, the WV spectrum of the optical phase data $\tilde{y}$ can be computed by Fourier transforming the above $C_{\tilde{y}\tilde{y}}(t,\tau)$ and the numerical results are shown in Fig.\ref{fig:spectrum of yy}. It should be noted that, since the WV spectrum can have negative values, we take the absolute value of the WV spectrum in the log-plot with different intra-cavity powers. In addition, Fig.\,\ref{fig:spectrum of xx and xdW} show the spectrum of $C_{xx}$ and $C_{xdW}$ terms in Eq.\,\eqref{eq:correlation of xx}. Besides the usual mechanical response peak at $\omega_m$, the distinct SN features emerges at around the SN frequency $\omega_q$.

The peak structure appearing in the above numerical results can be explained through qualitative analysis of Eq.~\eqref{eq:correlation of xx}. Before proceeding with this analysis, let us revisit the behavior of $h_1(t)$ from Section~\ref{section:Riccati equation}: (1) $h_1(t)$ converges to an equilibrium point $h_1^{\rm{eq}}$, which serves as a DC baseline; (2) spectral analysis reveals that $h_1(t)$ exhibits oscillatory behavior primarily at frequency $2\omega_q$, with secondary contributions at $4\omega_q$ in certain parameter regimes.

Now for simplicity, we'll expand the integrand of the term $\alpha^2C^m_{xx}(t,\tau)$ in Eq.~\eqref{eq:correlation of xx} as an example and focus on one term of it without loss of generality. The term we consider can be rewritten using the trigonometric identity:
\begin{equation}\label{eq:example term of Cxx}
\begin{split}
    &\int_0^{t-\frac{|\tau|}{2}}ds\ e^{\gamma_m(s-t)}h^2_1(s)\cos\omega_{mc}(s-t-\frac{\tau}{2})\cos\omega_{mc}(s-t+\frac{\tau}{2})\\
    &=\frac{e^{-\gamma_mt}}{2}\int_0^{t-\frac{|\tau|}{2}}ds\ e^{\gamma_ms}h^2_1(s)\left[\cos2\omega_{mc}(s-t)+\cos\omega_{mc}\tau\right],
\end{split}
\end{equation}

The structure of the above integral in the frequency domain\,(conjugate to $\tau$) can be analyzed using the following relationship:
\begin{equation}
    \int_{-\infty}^{+\infty}e^{-i\Omega\tau}d\tau \left[\int_{0}^{t-\frac{|\tau|}{2}}ds\ f(s)\right]=\frac{2\cos{2\Omega t}}{\Omega}{\rm Im}[\tilde{f}(2\Omega)],
\end{equation}
which relates the Fourier transform of the function $\int_{0}^{t-\frac{|\tau|}{2}}ds\ f(s)$ and $\tilde{f}(\Omega)$ as the Fourier transform of $f(s)$.  For example, suppose the function $f(s)$ oscillates at a main frequency $\omega$, then the spectrum\,(conjugate to $\tau$)  of $\int_{0}^{t-\frac{|\tau|}{2}}ds\ f(s)$ has a peak at $\Omega=\omega/2$.


Now we analyze the structure of Eq.~\eqref{eq:example term of Cxx}, shown in the upper panel of Fig.\,\ref{fig:spectrum of xx and xdW}. The first integrand $e^{\gamma_ms}h^2_1(s)\cos2\omega_{mc}(s-t)$ contains frequency components at $2\omega_{mc}$, $2\omega_q\pm2\omega_{mc}$, and $4\omega_q\pm2\omega_{mc}$ due to the beating between $h^2_1(s)$ and $\cos2\omega_{mc}(s-t)$. After integration over $s$, the frequency components of the resulting function are halved, yielding peaks at $\omega_m$, $\omega_q\pm\omega_m$, and $2\omega_q\pm\omega_m$. A similar analysis can show that the second term produces identical frequency components as the first term, and the WV spectrum of $\alpha^2C^{\rm th}_{xx}(t,\tau)$ peaks only at $\omega_m$.  Furthermore, the term $\alpha^2C^{m}_{xx}(t,\tau)$ dominates the spectrum and exhibits a clear oscillatory structure with a characteristic period of $\pi/t$.

{For the remaining term $\alpha C_{xdW}(t,\tau)$, we can demonstrate that $h_1(t-|\tau|/2)$ contains frequency components at half of those in $h_1(t)$:
\begin{equation}
    \int_{-\infty}^{+\infty}d\tau\ h_1(t-\frac{|\tau|}{2})e^{-i\Omega\tau}=2h_1(2\Omega)^*e^{-2i\Omega t}+2h_1(2\Omega)e^{i2\Omega t}.
\end{equation}
Implementing the same method, we can conclude that $\alpha C_{xdW}(t,\tau)$ exhibits identical frequency components to those found in $\alpha^2C^m_{xx}(t,\tau)$.

Combing all the terms above, the peak structure of the WV spectrum of $\tilde{y}(t)$ in Fig~\ref{fig:spectrum of yy} can also be explained. These peak structures around $\omega_q$ demonstrate that the WV spectrum of CCSN theory exhibits significantly distinguishable features compared to QG theory.  In our simulations, the distinction between the two theories becomes most pronounced when the intra-cavity power is $P_{\textrm{cav}}\sim1\mu$W. Focusing specifically on the frequency components at $\omega_q\pm\omega_m$, the relative difference between CCSN and QG theories can reach approximately 10\,dB, which represents a substantial improvement compared to the stationary case analyzed previously. Therefore, the subsequent discussion will be conducted using these parameters.

\begin{figure*}
    \centering
    \includegraphics[scale=0.154]{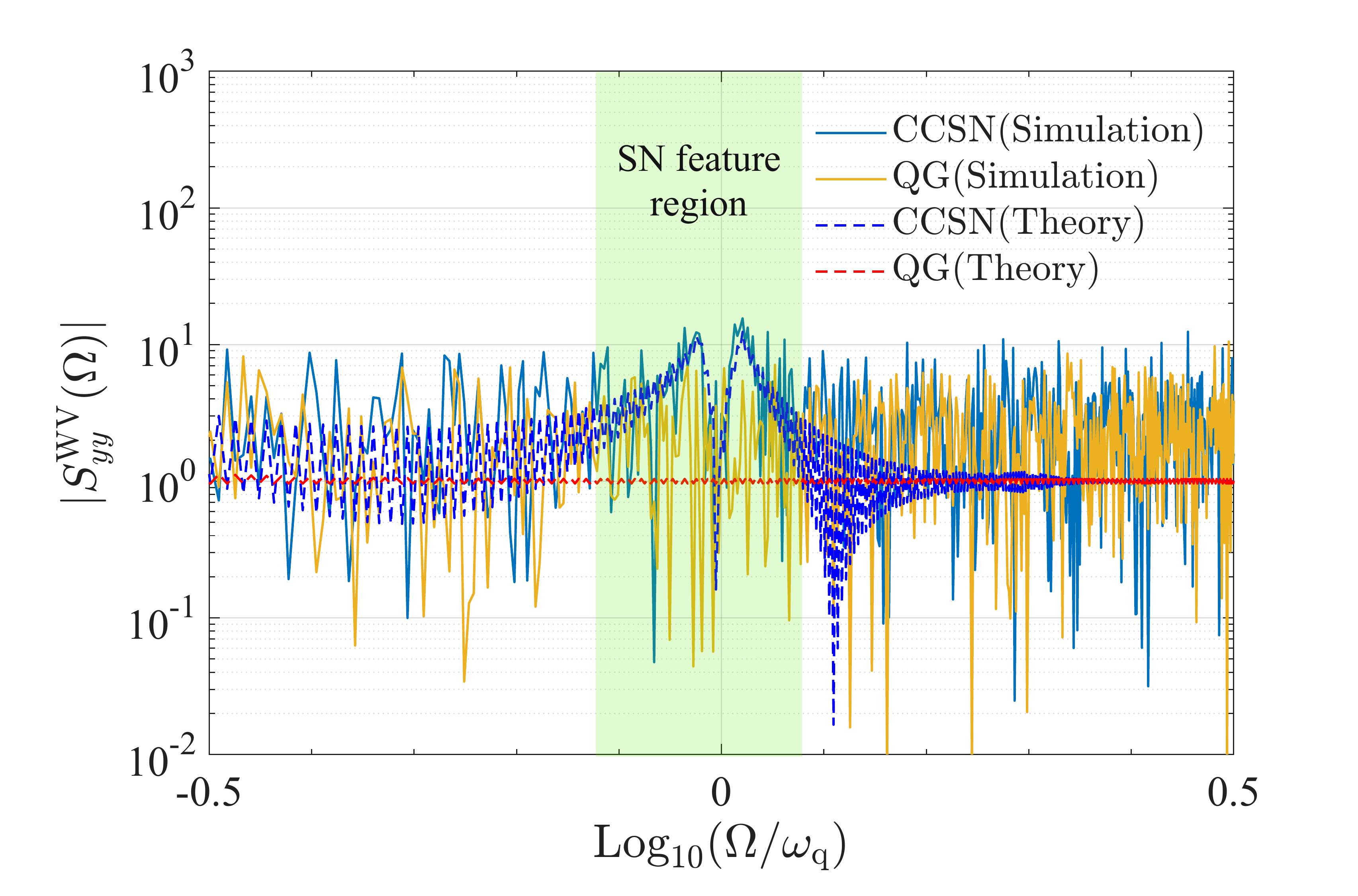}
    \includegraphics[scale=0.154]{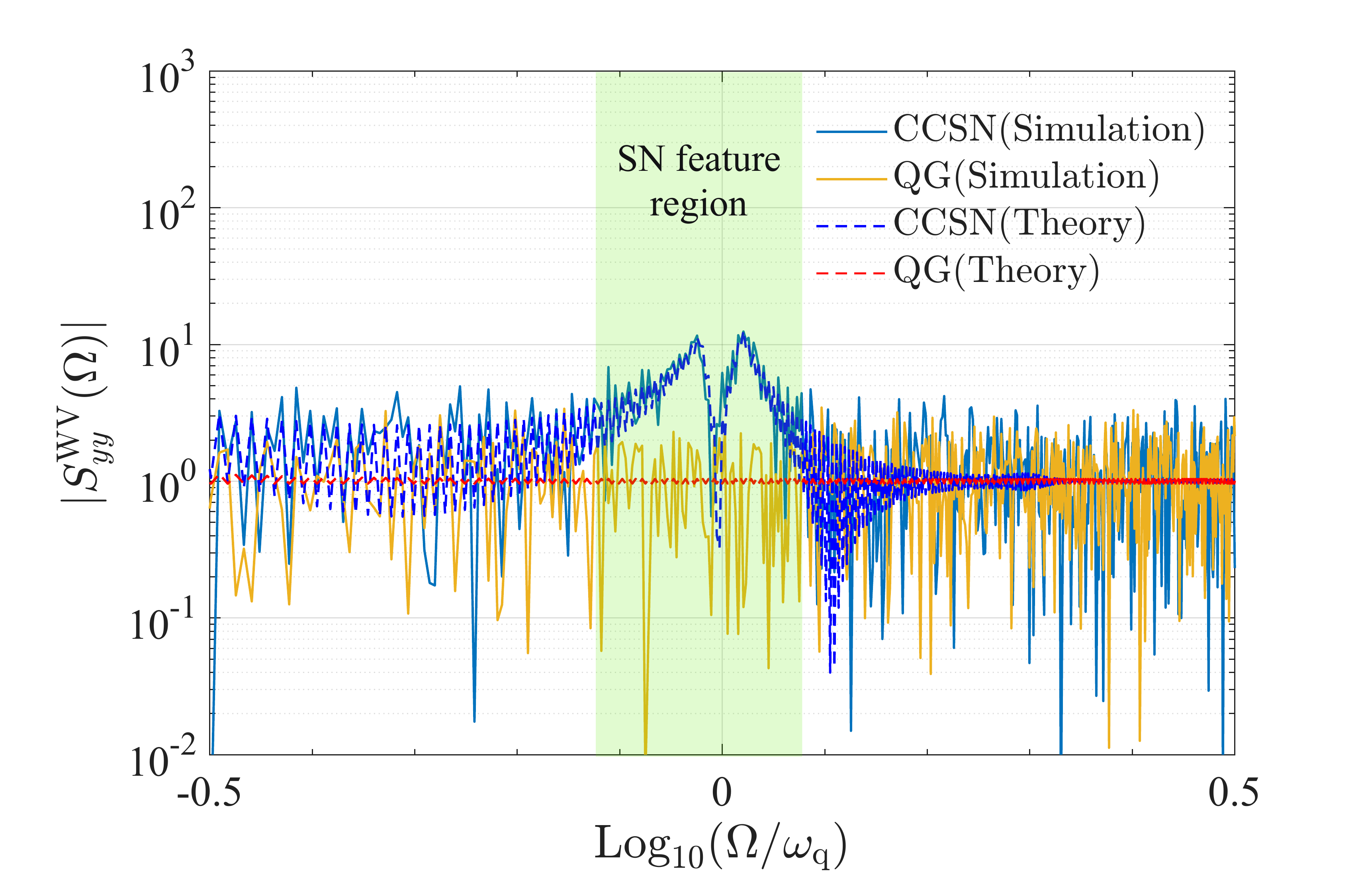}
    \includegraphics[scale=0.154]{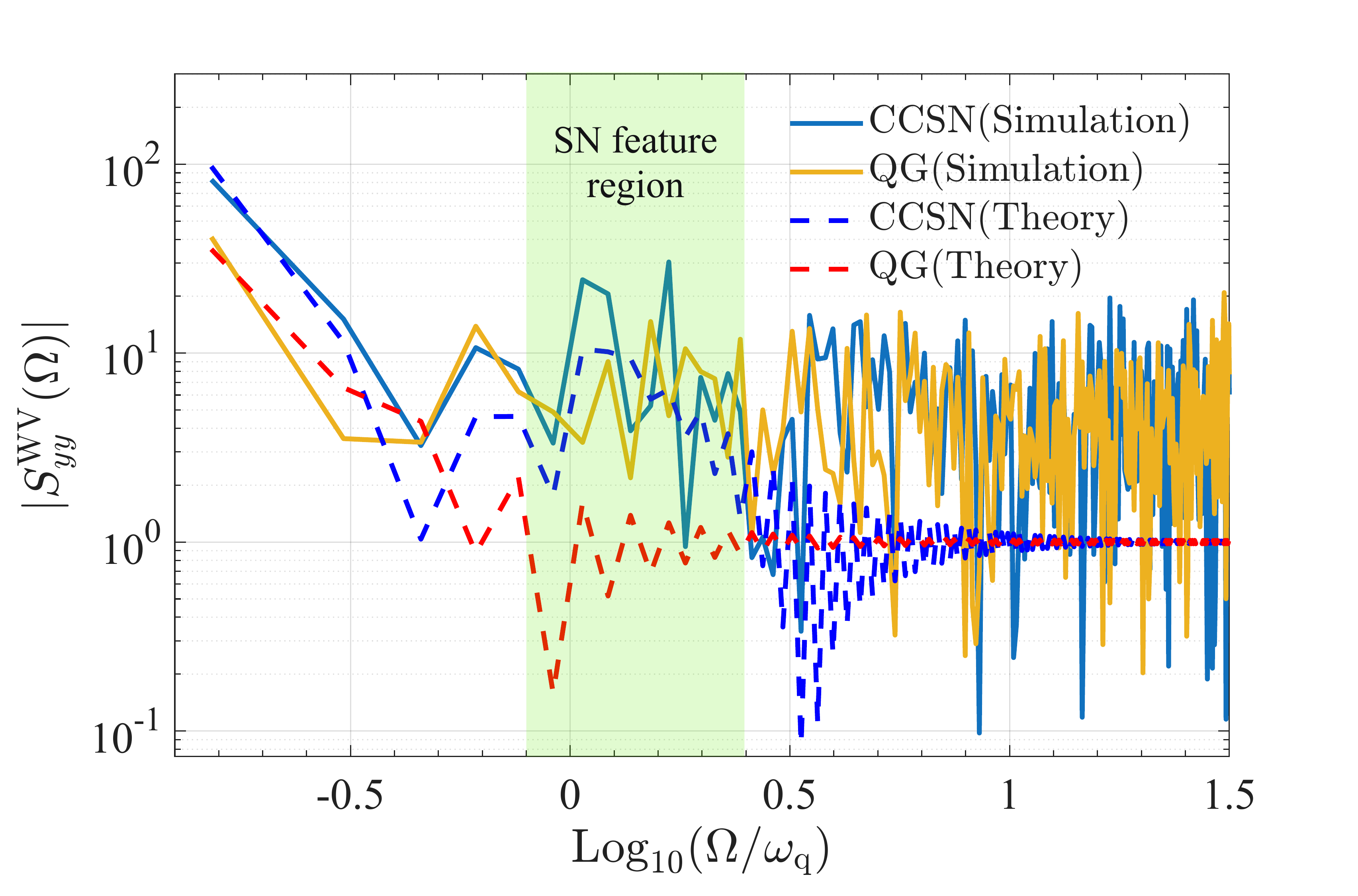}
    \includegraphics[scale=0.154]{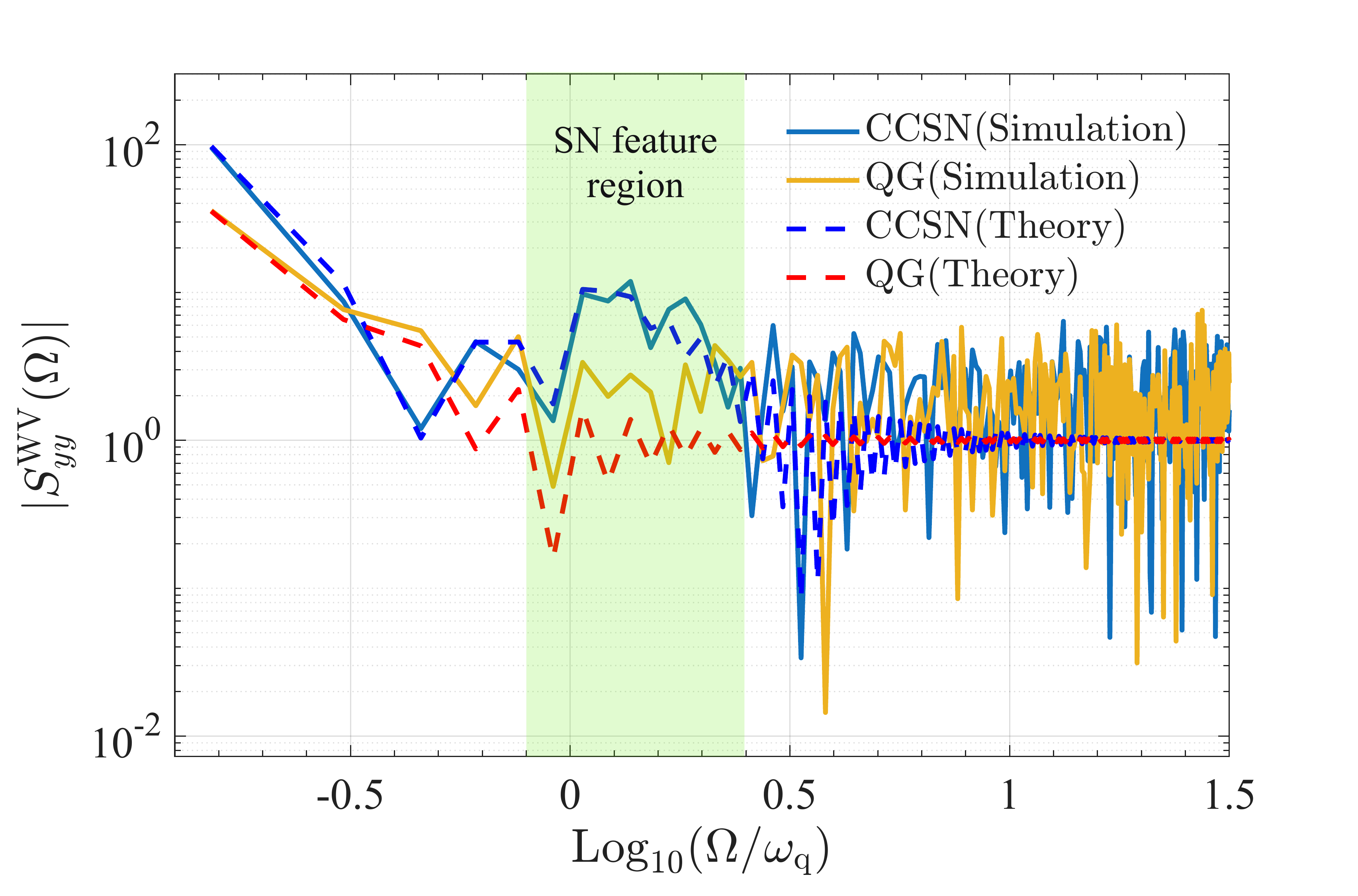}
    \caption{Mock-data simulation of the WV spectrum with different measurement time and different numbers of experimental trials. Upper panel: the spectrum with $T_{\rm obs}=10^3$\,sec obtained by averaging over $10^4$ independent trial\,(left panel) shows that the distinct feature between SN gravity\,(blue curve) and QG\,(yellow curve) is obscured by the fluctuations. While with $10^5$ trials,(right panel), the statistical fluctuations are significantly reduced, resulting in a clearer distinction between the two theories. Lower panel: When $T_{\rm obs}=40$\,sec, the SN feature can be hardly distinguished with $10^2$ trials\,(left panel), but it becomes distinguishable with $10^3$ trails\,(right panel).}
    \label{fig:stochastic simulations}
\end{figure*}
\subsection{Mock data analysis}

\begin{figure*}
    \centering
    \includegraphics[scale=0.24]{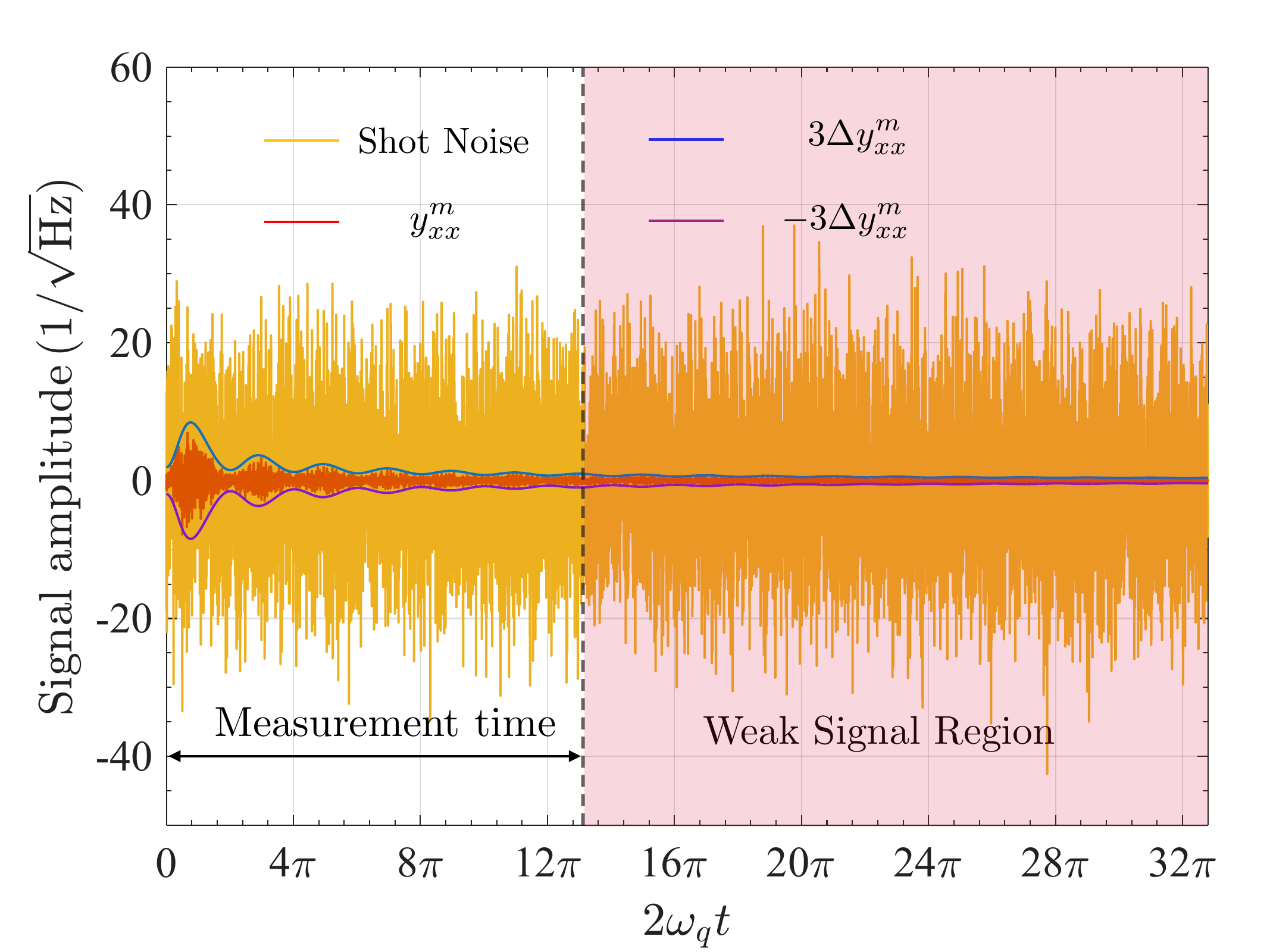}
    \includegraphics[scale=0.24]{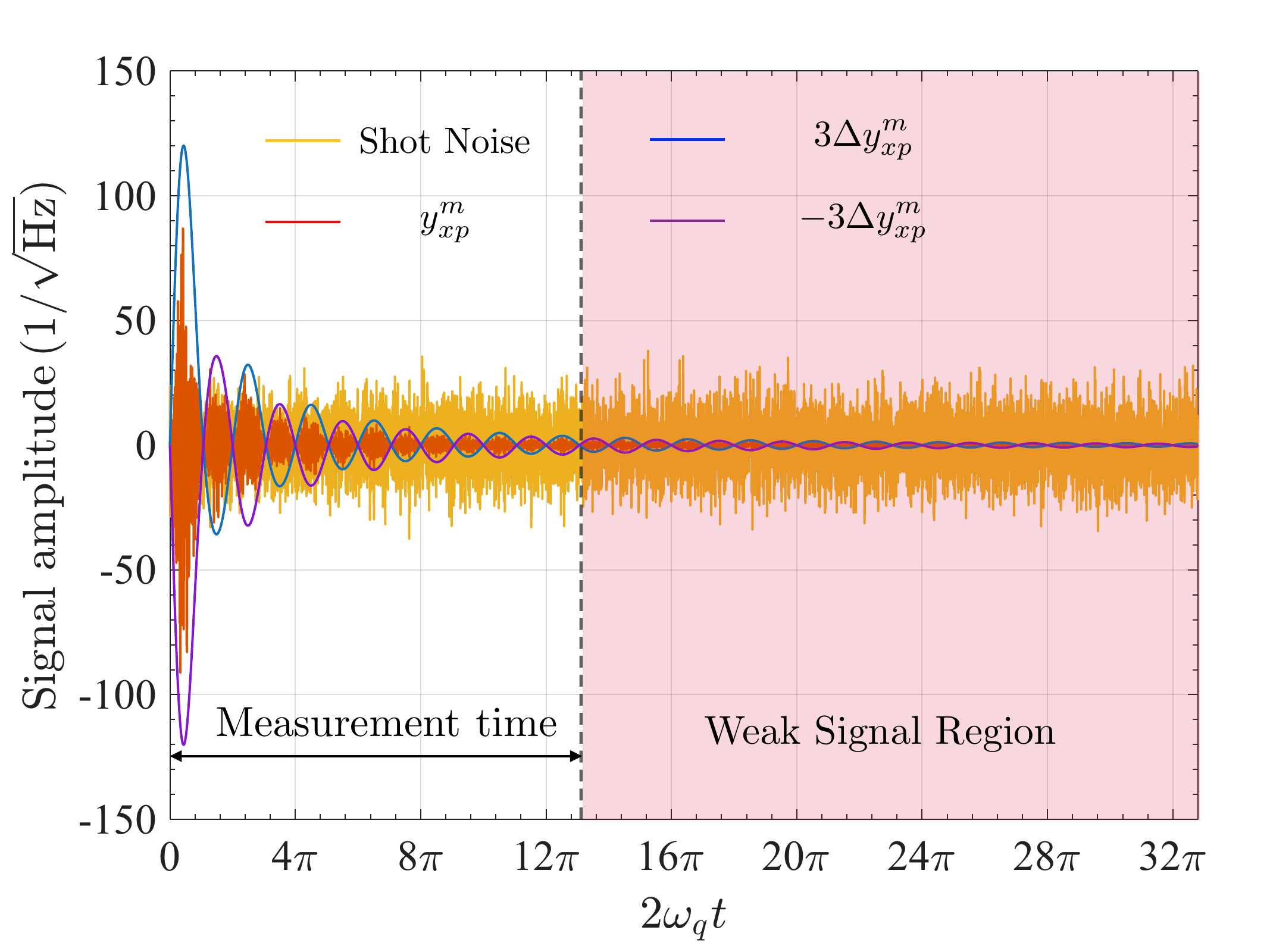}
    \caption{A comparison of the stochastic signal containing SN feature $y_{xx}^m(t)$ and $y_{xp}^m(t)$ to the shot noise. The $\Delta y_{xx}^m(t)$ and $\Delta y_{xp}^m(t)$ represent the variance of signal. The envelope of the stochastic signals contain the SN feature, which is mainly contributed in the first 40 sec.}
    \label{fig:SN signal to shot noise}
\end{figure*}
The above results represent an idealized scenario that assumes an infinite number of experimental realizations for statistical averaging. In practical experiments, we can only perform a finite number of measurements, which inevitably introduces fluctuations in the observed WV spectrum, hence obscuring the signature of CCSN by stochastic fluctuations. As an example, the upper panel of Figure~\ref{fig:stochastic simulations} illustrates the WV spectrum obtained by averaging over $10^4$ and $10^5$ independent experimental repetitions, respectively. The figure shows that the emergence of the SN feature around $\omega_q$ requires $10^5$ experimental trials, which is experimentally inefficient.

In fact, experiments with short measurement time are more conducive to distinguishing the two models. For instance, if we decrease the measurement time to $40$\,sec, then the distinguishable SN feature can emerge in the spectrum with only $10^3$ independent experiments trials, as shown in the lower panel of Fig.\,\ref{fig:stochastic simulations}. The underline reason is that although longer measurement time improves the frequency resolution and enhances the peak structure, it however introduces more shot noise in one experimental repetition.

To illustrate the above argument, we plot the comparison between the signal containing SN features $\alpha x^m(t)$ and the shot noise in Fig.~\ref{fig:SN signal to shot noise} from the output quadrature data Eq.\,\eqref{eq:evolution of x(classical noise)}, where we define
\begin{equation}
    y_{xx}^m(t)=\sqrt{2}\alpha^2V_{xx}(t)\frac{dW}{dt},\ y_{xp}^m(t)=\sqrt{2}\alpha^2\frac{V_{xp}(t)}{M\omega_{mc}}\frac{dW}{dt}.
\end{equation}
 It is clearly shown in the figure that only in the early stage of non-stationary evolution does the signal containing SN features with an intensity matching the shot noise, while the subsequent time was mainly dominated by the shot noise. In the following discussion, we all assumes a $T_{\rm obs}=40$\,sec measurement time.

To have a more realistic estimation, a mock data simulation is performed in this section, targeted at finding the minimum number of experimental repetitions needed to reliably distinguish between the predictions of SN gravity and QG by using the WV spectrum.


Let us first characterize the statistical fluctuations in the WV spectrum. For each experimental realization, we can define
\be
S_{y_iy_i}(\Omega)=\int^\infty_{-\infty}\frac{d\tau}{2\pi}e^{-i\Omega\tau}\left[y_i\left(t+\frac{\tau}{2}\right)y_i^*\left(t-\frac{\tau}{2}\right)\right],
\ee
and the mean WV spectrum over $N$-realizations
\be
\langle S^{\rm{WV}}_{\tilde{y}\tilde{y}}(t,\Omega)\rangle=\frac{1}{N}\sum_{i=1}^NS_{y_iy_i}(t,\Omega).
\ee
The variance of the WV spectrum is defined as
\begin{equation}
    \sigma^2_{\tilde{y}\tilde{y}}(t,\Omega)={\rm lim}_{N\rightarrow \infty}\left[\frac{1}{N}\sum_{i=1}^NS^2_{y_iy_i}(t,\Omega)-\langle S^{\rm{WV}}_{\tilde{y}\tilde{y}}(t,\Omega)\rangle^2\right].
\end{equation} 
According to the central limit theorem, the above mean WV spectrum over $N$-realizations satisfies the normal distribution: $\mathcal{N}\left[\mu_{\tilde y\tilde y},\sigma^2_{\tilde{y}\tilde{y}}(t,\Omega)/N\right]$
where $\mu_{\tilde y\tilde y}={\rm lim}_{N\rightarrow \infty}\langle S^{\rm{WV}}_{\tilde{y}\tilde{y}}(t,\Omega)\rangle$.
Moreover,  the absolute value of the averaged WV spectrum $x=|\langle S^{\rm{WV}}_{\tilde{y}\tilde{y}}(t,\Omega)\rangle|$ satisfies the folded normal distribution:
\begin{equation}
    f(x) = \sqrt{\frac{2}{\pi\sigma_{\tilde{y}\tilde{y}}^2}}\exp\left(-\frac{x^2+\mu_{\tilde{y}\tilde{y}}^2}{2\sigma_{\tilde{y}\tilde{y}}^2}\right)\cosh\left(\frac{\mu_{\tilde{y}\tilde{y}} x}{\sigma_{\tilde{y}\tilde{y}}^2}\right),\quad (x>0).
\end{equation}

\begin{figure}[h]
    \centering
    \includegraphics[scale=0.45]{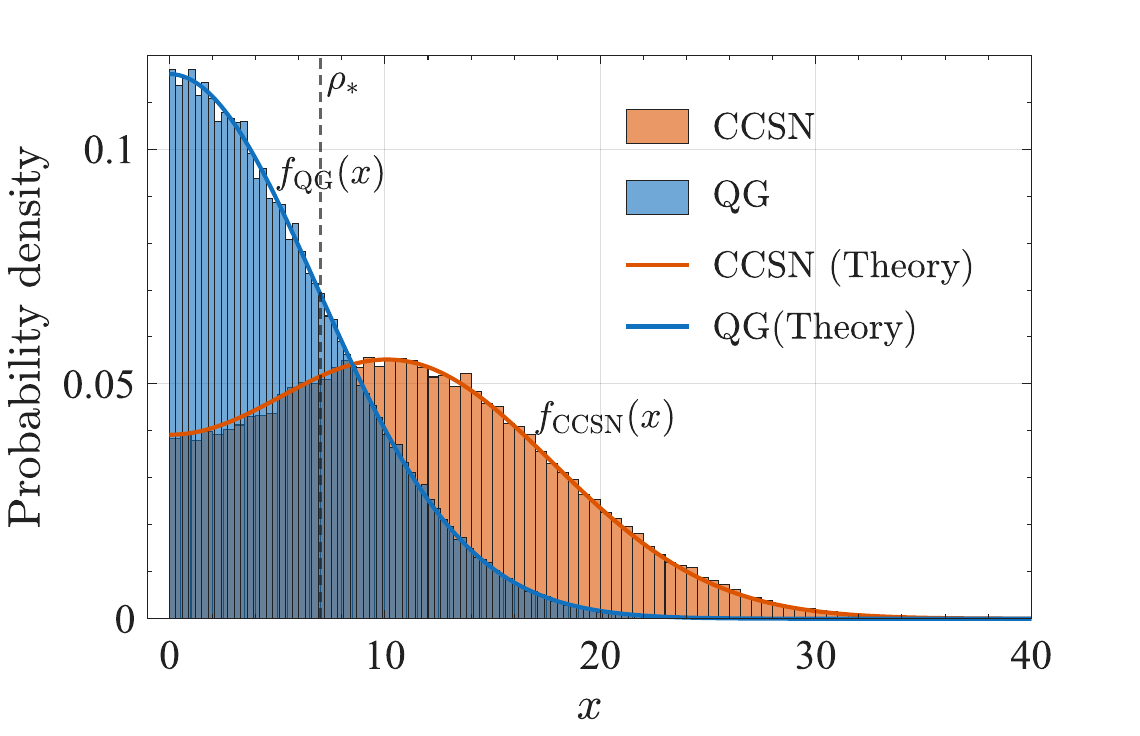}
    \includegraphics[scale=0.45]{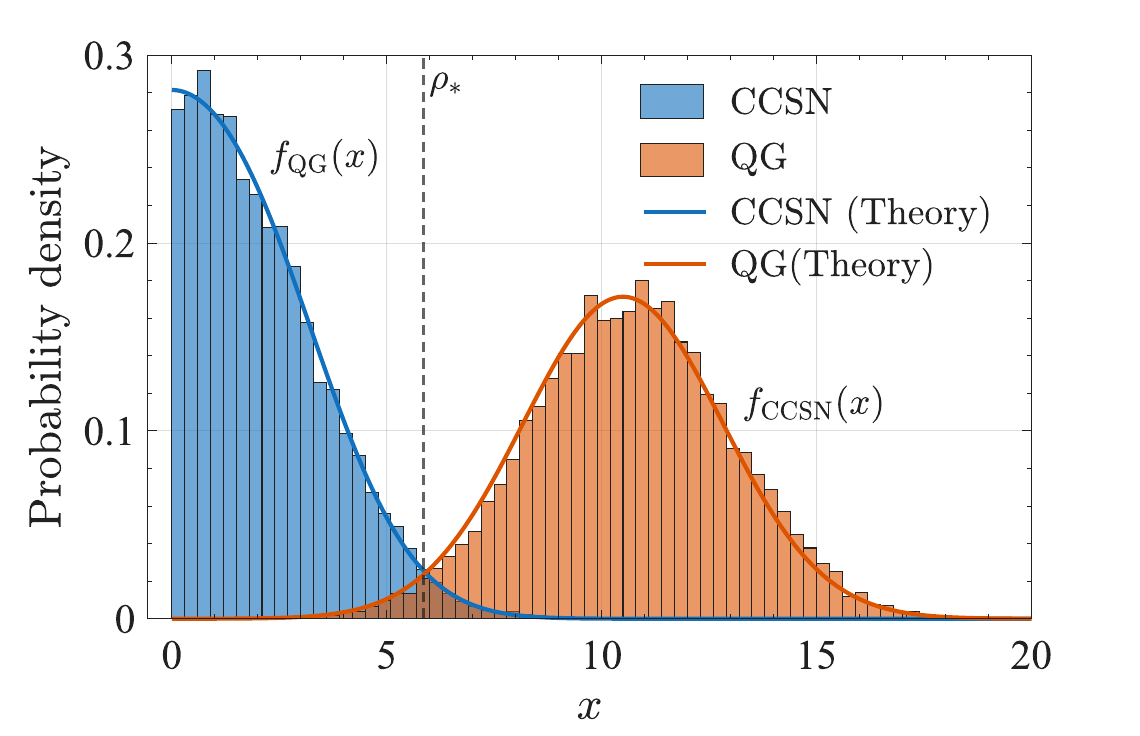}
    \caption{Upper panel: Probability density distribution of the averaged WV spectrum at frequency $\omega_q+\omega_m$ with $10^2$ repeated experiments. The histogram shows the simulation results while the solid curves represent the theoretical predictions for both standard quantum mechanics/,(red) and SN gravity/,(blue). The vertical dashed line indicates the optimal threshold value $\rho_*=7$ that yields equal error rates of $\mathcal{F}=\mathcal{D}=0.3$. Lower panel: With $10^3$ repeated experiments, $\rho_*=5.85$ and error rates decrease dramatically to $\mathcal{F}=0.023$.}
    \label{fig:distribution}
\end{figure}

In Fig.\,\ref{fig:distribution}, we plot the distribution functions $f_{\rm SN}(x)$ and $f_{\rm QG}(x)$, together with the histogram from our Mock-Data simulation for the averaged WV spectrum in case of $10^2$ and $10^3$ experimental realizations. For quantitatively distinguishing the SN theory and QG theory, we set a threshold value $\rho^*$, so that the SN model is favored for measurement results exceeding $x_{\rm exp}>\rho_*$; otherwise, we accept the QG model.  Subsequently, we can define two important error metrics: the false alarm rate $\mathcal{F}$ and the false dismissal rate $\mathcal{D}$. The false alarm rate represents scenarios that the $x_{\rm exp}>\rho_*$ while QG is the actual underlying physics, leading to an incorrect judgement of SN model. The false dismissal rate quantifies the scenarios that $x_{\rm exp}<\rho_*$ while CCSN is the reality, causing an erroneous conclusion of the QG model.  Mathematically, these two error metrics can be formulated as:
\begin{equation}
    \mathcal{F}=\int_{\rho_*}^{\infty}f_{\rm{QG}}(x)dx,\quad \mathcal{D}=\int_{-\infty}^{\rho_*}f_{\rm{SN}}(x)dx,
\end{equation}
which are required to be small for a clear distinction between SN and QG theories.

Suppose a balanced decision criterion is taken so that we can derive a critical $\rho_*$ when we take $\mathcal{F}=\mathcal{D}$.  Focusing on the distribution at the characteristic peak $\omega_q+\omega_m$, our theoretical analysis of mean and variance indicates that the optimal threshold for $10^2$ and $10^3$ repeated experiments are $\rho_*= 7$ and $\rho_*=5.85$, and the false alarm rate and false dismissal rate in this case are $\mathcal{F}=\mathcal{D}=0.3$ and $0.023$, respectively.
However, while the WV spectrum provides an alternative theoretical means to distinguish between SN gravity and standard QM, practical implementation requires numerous experimental repetitions to achieve statistical significance—an approach that proves practically difficult and resource-sensitive. For practical experimental implementation, a more efficient approach is needed to leverage the full covariance matrix structure for statistical inference, significantly reducing the number of required measurements while preserving robust discriminative power. Meanwhile, WV spectrum analysis here can help identify optimal parameters that maximize distinguishability between the SN and QG models in non-stationary processes.

\section{Statistical inference}\label{section:statistical inference}


The Wigner-Ville spectrum discussed in the previous section captures only partial information about a nonstationary Gaussian stochastic process. Extracting more complete information from such a process can help relax the requirement for a large number of experimental repetitions. By treating all time-series data points as components of a high-dimensional Gaussian random variable, we can construct the corresponding high-dimensional covariance matrix, which encodes significantly more information than the Wigner-Ville spectrum alone. This forms the foundation of our statistical inference approach.


\subsection{Covariance matrix}
Assuming the Gaussian random variables $\mathbf{Y}^i=\{Y_{t_1},Y_{t_2},...,Y_{t_n}\}^i$\,(with $i=\{\rm{QG},\rm{SN}\}$ represents two different gravity models) are the discretized mock-data generated by these two gravity models, which satisfies a high-dimensional random distribution $\mathbf{Y}^i \sim \mathcal{N}_n(\mu,\bm{\Sigma}^i)$.  The mean $\mu$ is zero for both two models, and the elements of covariance matrix $\bm{\Sigma}^i$ is defined as:
\begin{equation}\label{eq:covarice matrix}
    (\bm{\Sigma}^i)_{j k}=\mathbb{E}[Y_{t_j}Y_{t_k}]-\mathbb{E}[Y_{t_j}]\mathbb{E}[Y_{t_k}]=\mathbb{E}[Y_{t_j}Y_{t_k}].
\end{equation}
To obtain the concrete form of the covariance matrix elements, we substitute~\eqref{eq:measurement result} into~\eqref{eq:covarice matrix} and obtain: 
\begin{equation}
\begin{split}
        \mathbb{E}[{\rm{Y}}_{t_j}{\rm{Y}}_{t_k}] =&\alpha^2[(\bm{\Sigma}^i_{xx})_{jk}^m+(\bm{\Sigma}^i_{xx})_{jk}^{\rm th}]+\alpha({\bm{\Sigma}^i_{xdW})_{jk}}\\
        &+\frac{1}{2}\delta(t_j-t_k),
\end{split}
\end{equation}
where
\begin{widetext}
\begin{equation}
\begin{split}
    &\alpha^2(\bm{\Sigma}^i_{xx})_{jk}^m=\frac{\Lambda_*^4}{2\omega_q^2}\int^{\min(t_j,t_k)}_0ds\ e^{\frac{\gamma_m}{2}(2s-t_j-t_k)}\left(\frac{\omega_m}{\omega_{mc}}y_1(s)\cos{[\omega_{mc}(s-t_j)-\phi]}-\frac{2\omega_q}{\omega_{mc}}y_2(s)\sin{[\omega_{mc}(s-t_j)]}\right)\times[t_i\rightarrow t_j],\\,
\end{split}
\end{equation}
\begin{equation}
\begin{split}
    \alpha^2(\bm{\Sigma}^i_{xx})_{jk}^{\rm th}=\Lambda_*^2\frac{ \omega_m \gamma_m}{2 \omega_{mc}^2}\coth\left( \frac{\hbar \omega_m}{2 k_B T} \right)&\left[e^{-\frac{\gamma_m}{2}|t_j-t_k|}\left(\frac{\cos[\omega_{mc}(t_j-t_k)]}{\gamma_m}+\frac{\sin[\omega_{mc}|t_j-t_k|+\phi]}{2\omega_m}\right)\right.\\
    &\left.-e^{-\frac{\gamma_m}{2}(t_j+t_k)}\left(\frac{\cos[\omega_{mc}(t_j-t_k)]}{\gamma_m}+\frac{\sin[\omega_{mc}(t_j+t_k)+\phi]}{2 \omega_{m}}\right)\right],
\end{split}
\end{equation}
\begin{equation}
\begin{split}
\alpha{(\bm{\Sigma}^i_{xdW})_{jk}}=\frac{\Lambda_*^2}{2\omega_q} e^{-\frac{\gamma_m}{2}|t_j-t_k|}\left[\frac{\omega_m}{\omega_{mc}}y_1(\min(t_j,t_k))\cos{(\omega_{mc}|t_k-t_j|+\phi)}+\frac{\omega_q}{\omega_{mc}}y_2(\min(t_j,t_k))\sin{(\omega_{mc}|t_k-t_j|)}\right].
\end{split}
\end{equation}
\end{widetext}
It is important to note that due to the discretized sampling of the continuous process, the Wiener increment $dW/dt$ is not an ideal delta function, but rather a Gaussian random variable with zero mean and variance $1/\Delta t$, where $\Delta t$ is the time-sampling resolution. Therefore, we have the following property of the discretized Wiener increment:
\begin{equation}
    \mathbb{E}\left[\Delta W(t_j)\Delta W(t_k)\right]= \Delta t \delta_{jk},
\end{equation}
where the continuous delta function $\delta(t_j-t_k)$ in our formulation is replaced by $\delta_{jk}/\Delta t$ in the discrete case, and the $\delta_{jk}$ is the Kronecker delta. The Gaussian probability density function\,(PDF) that constructed from this high-dimensional random variable $\mathbf{Y}^i$ is:
\begin{equation}
f_i(\mathbf{Y}^i)= \frac{1}{\sqrt{{(2\pi)^n/|\Sigma^i|}}}\exp(-\frac{1}{2}[\mathbf{Y}^i]^{T}(\Sigma^{i})^{-1}\mathbf{Y}^i).
\end{equation}

\subsection{Dimensionality reduction and error rate}
The next critical step is to determine which gravity model\,(QG or SN) the experimental data aligns with, based on the covariance matrices derived for each model. This involves statistically comparing the covariance structure of the experimentally observed data with the theoretical predictions of both models.  A naive way is to generate the data $Y_{t_i}$ from the two gravity models and define a new stochastic quantity $\mathcal{Y}_{ij}=Y_{t_i}Y_{t_j}$, and then we have:
\be
\bar{\mathcal{Y}}_{ij}=\mathbb{E}[Y_{t_i}Y_{t_j}],
\ee
and
\begin{equation}
\begin{split}
    {\rm Var}(\mathcal{Y}_{ij})=\mathbb{E}[Y_{t_j}^2]\mathbb{E}[Y_{t_k}^2]+\mathbb{E}[Y_{t_j}Y_{t_k}]^2,
\end{split}
\end{equation}
in deriving which we have used the Wick theorem to factorize the four-point correlation function to the second-point correlations. By applying the Cauchy-Schwarz inequality, we obtain:
\begin{equation}
    \frac{\bar{\mathcal{Y}}_{ij}}{\sqrt{\mathrm{Var}[\mathcal{Y}_{ij}]}} = \frac{\mathbb{E}[Y_{t_j}Y_{t_k}]}{\sqrt{\mathbb{E}[Y_{t_j}^2]\mathbb{E}[Y_{t_k}^2]+\mathbb{E}[Y_{t_j}Y_{t_k}]^2}} \leq \frac{1}{\sqrt{2}}.
\end{equation}
Taking the equality case as our baseline, we can estimate the minimum number of required experiments. Suppose the experiment is repeated independently for $N$ times and we construct a statistical quantity $\langle \mathcal{Y}\rangle_{N}=\sum_{n=1}^N\mathcal{Y}^{(n)}_{ij}/N$, the central limit theorem tells us that for sufficiently large $N$, the variance of $\langle \mathcal{Y}\rangle_{N}$ reduces to $\mathrm{Var}(\mathcal{Y}_{ij})/N$. If we require the error in the covariance matrix to be less than $10\%$ of $\bar{\mathcal{Y}}_{ij}$,  the $3\sigma$ statistical significance leads to the equation for the required experiment repitition $N$:
\begin{equation}
    3\sqrt{\frac{\mathrm{Var}(\mathcal{Y}_{ij})}{N}} \leq 0.1\Sigma_{ij},
\end{equation}
the solution of which show that the minimal number of experiments required is $N=1800$, which is practically resources-inefficient.
\begin{figure}[h]
    \centering
    \includegraphics[scale=0.5]{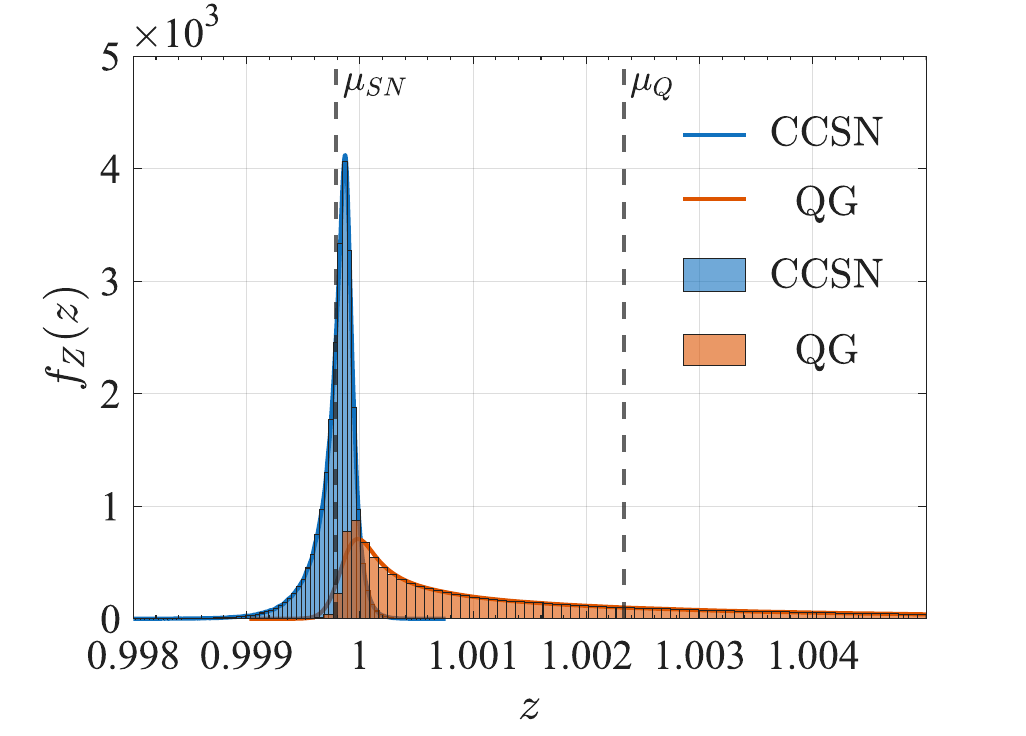}
    \includegraphics[scale=0.5]{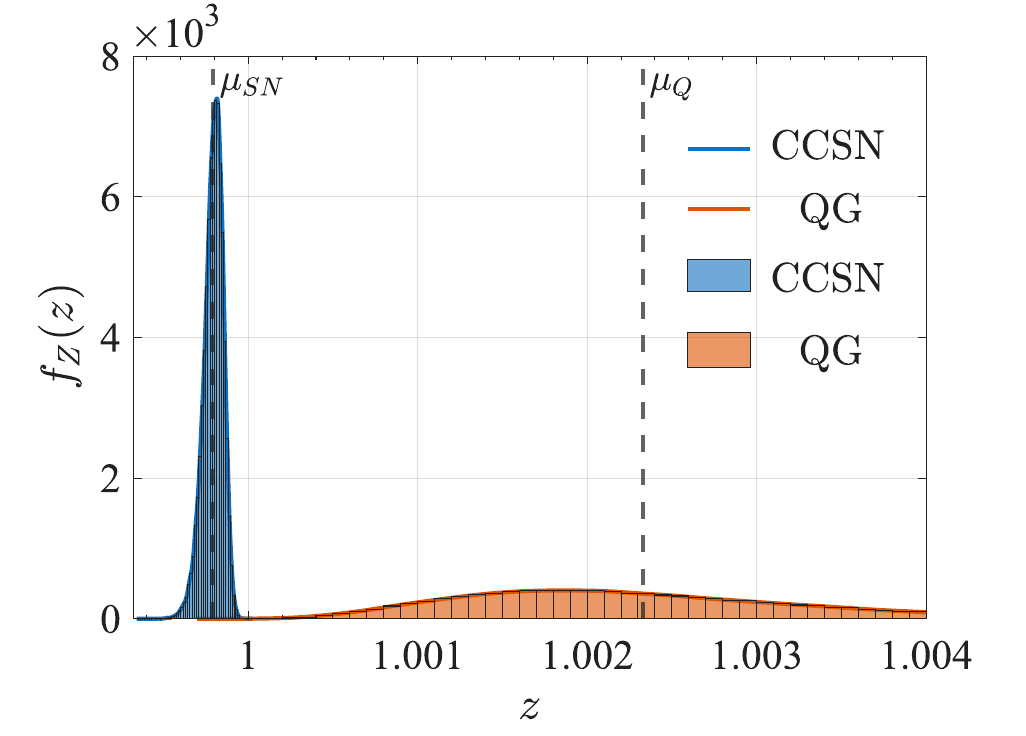}
    \caption{Probability density distributions of the log-likelihood ratio. Upper panel: Single-dataset distributions for CCSN theory\,(red) and standard quantum mechanics\,(blue) with 4.5 dB of squeezing. Lower panel: Distributions after averaging over ten independent experimental datasets, showing significantly improved separation between the two theoretical models.}
    \label{fig:PDF of z}
\end{figure}

\begin{figure}[h]
    \centering
    \includegraphics[scale=0.5]{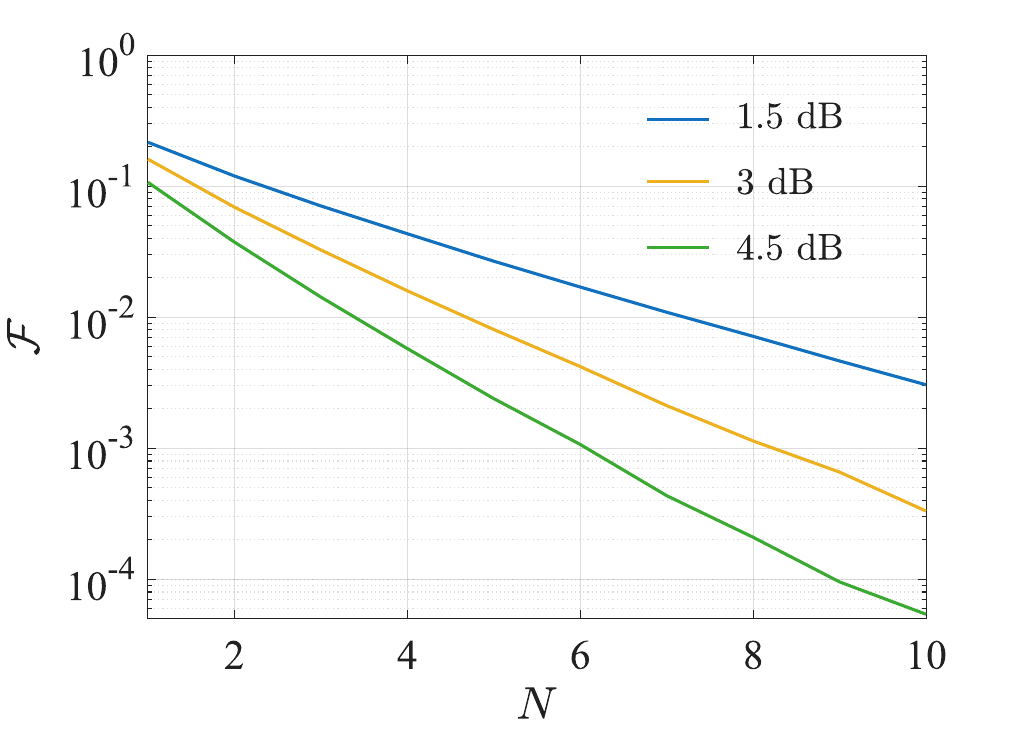}
    \includegraphics[scale=0.5]{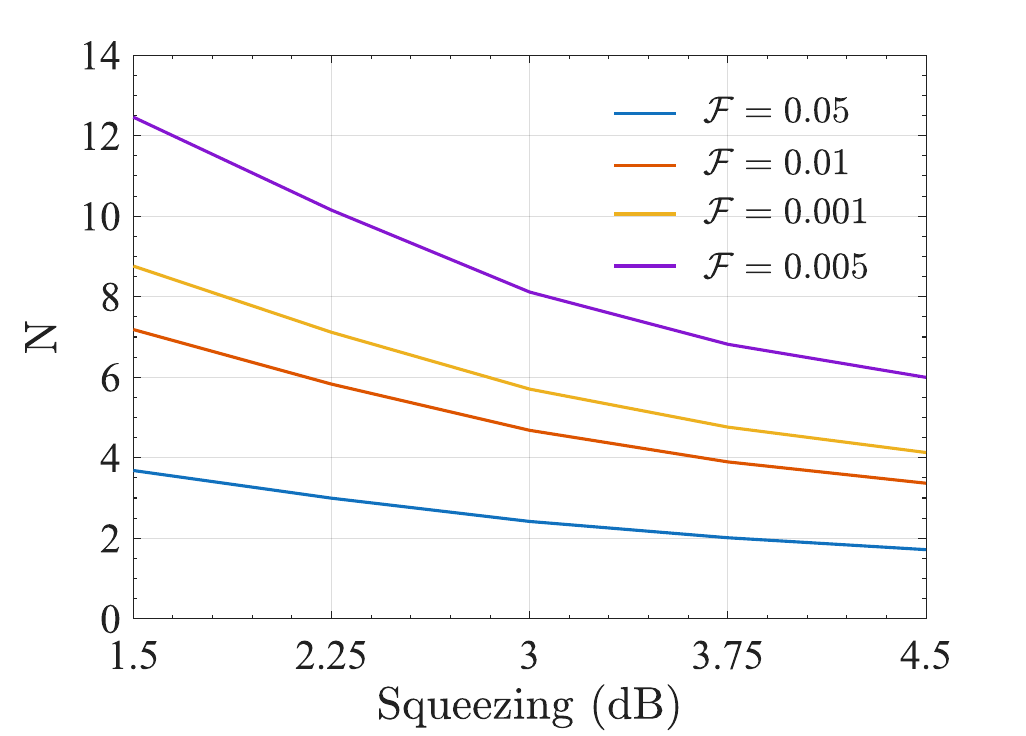}
    \caption{Upper panel: Error rates for the statistical inference method as a function of experimental trial numbers. Lower panel: Number of required experimental repetitions to achieve equal error rates at different squeezing levels.}
    \label{fig:error rate}
\end{figure}

In fact, we do not need to accurately estimate all covariance matrix elements. We can use the dimension-reduction\,\cite{van2009dimensionality,10.5555/1076819,1183897} approach to extract the distinctive features between two gravity models. Essentially, we can transform the high-dimensional random variable $\mathbf{Y}^i$ into a one-dimensional random variable $Z^i$ while preserving the characteristics required for model discrimination. Following this principle, we construct the following log-likelihood ratio of the two competing gravity models:
\begin{equation}
    Z^i(\mathbf{Y}^i) = \frac{\log{f_{\rm{SN}}(\mathbf{Y}^i)}}{\log{f_{\rm{QG}}(\mathbf{Y}^i)}},
\end{equation}
which is also a stochastic quantity.  For data generated according to one gravity model, the distribution of $Z^i$ will differ significantly between CCSN theory and QG/,(or standard QM).

Based on the multivariate normal distribution of both gravity models, we can employ Monte Carlo methods to generate the probability distribution function of the likelihood ratio $Z^i(\mathbf{Y}_i)$. In this study, we generated one million independent simulated datasets for each gravity model. For each dataset, we calculated the log-likelihood ratio $Z^i$ and subsequently obtained the probability density function from the histogram using kernel density estimation, which is shown in the upper panel of Fig.\,\ref{fig:PDF of z}. 

Similar to our analysis of the WV spectrum, we can use the above PDF to calculate error rates for the statistical inference method. The false alarm rate $\mathcal{F}_Z$ and false dismissal rate $\mathcal{D}_Z$ can be expressed as:
\begin{equation}
    \mathcal{F}_Z=\int_{-\infty}^{\rho_*}f^Z_{\rm{QG}}(z)dz,\quad \mathcal{D}_Z=\int_{\rho_*}^{\infty}f^Z_{\rm{SN}}(z)dz,
\end{equation}
where we also set $\mathcal{F}_Z=\mathcal{D}_Z$ to establish a balanced decision criterion that determines the optimal threshold. As shown in Fig.~\ref{fig:PDF of z}, suppose one experiment data series is performed, the error rate is unacceptably high with $\mathcal{F}_Z=0.107$ while the  $f^Z_{\rm{SN}}(z)$ and $f^Z_{\rm{QG}}(z)$ already exhibit differences. This suggests that averaging over repeated experiments could reduce the error rate. Suppose the experiment is independently repeated for $N$-times, we can construct an averaged random variable $\bar{Z}^i=\sum_{n=1}^{N}(Z^i)_n/N$. The probability density function of this averaged random variable can also be computed by using Monte Carlo sampling. The lower panel of Fig.~\ref{fig:PDF of z} illustrates the distributions when $N=10$, which shows a substantially decreased variance, leading to larger separation between the two theoretical models, and the error rates decrease to $5.4\times10^{-5}$.

We further investigated the scaling behavior of error rates as a function of experimental repetitions for different initial states, including the minimum number of repetitions required to achieve consistent error rates across these states. The different squeezed thermal states are generated by setting the minimal quadrature variance to $10^3$ times the zero-point fluctuation level, while varying the maximized quadrature variance. As shown in Fig.\,\ref{fig:error rate}, our quantitative analysis demonstrates the remarkable efficiency of our approach. Notably, with only modest squeezing\,(1.5 dB), a mere 10 experimental repetitions suffice to suppress error rates below 1\%— representing a significant improvement in resource efficiency compared to the Wigner-Ville spectrum method.

\section{Summary and discussion}\label{section:summary and discussion}
In conclusion, we have systematically investigated the non-stationary dynamics of an optomechanical system under the influence of SN gravity, taking into account the effect of continuous quantum measurement. The key signature that distinct SN gravity from quantum gravity comes from the SN evolution of the uncertainty ellipse—the cross-section of the Wigner function of the test mass, with frequency $2\omega_q$.
During the non-stationary evolution, the measurement process continuously extracts position information from the test mass, causing the uncertainty ellipse to gradually converge toward a steady state.  When the measurement strength is high, the test mass position information can be extracted on timescales much shorter than the ellipse's rotation period, leads to a rapid convergence prevents us from extracting ellipse's rotation dynamics from the optical output. Therefore, a sufficiently weak measurement strength is essential to preserve the rotational signatures. Additionally, since the rotational information manifests in the oscillatory behavior of the second-order moments, achieving a detectable signal requires an initial state with sufficient "squeezing" of the error ellipse.

The SN signature in the non-stationary optical output can show in the Wigner-Ville\,(WV) spectrum. By calculating the WV spectrum of the output light, we can identify characteristic peaks at frequencies $\omega_q \pm \omega_m$, distributed around the SN frequency $\omega_q$. This spectral signature could serve as a clear indicator for comparing the relative differences between SN and quantum gravity models.

For an accurate resolution of the featured SN signals in the WV spectrum, the measurement time must exceed the oscillation period of the mechanical oscillator—requiring sufficiently fine frequency resolution. For a mHz mechanical oscillator, this corresponds to measurement durations of approximately $T_{\rm obs}=10^3$\,sec. Under these conditions, the WV spectrum can theoretically reveal clear additional peak structures at $\omega_q \pm \omega_m$ that are unique to CCSN theory, achieving relative differences of approximately 10\,dB compared to QG. However, practical measurements suffer from random fluctuations that severely contaminate these peak structures.  Only by averaging over numerous experimental repetitions can we adequately suppress shot noise fluctuations—requiring approximately $10^5$ experimental trials to effectively distinguish between the two models.  However, if our objective is to distinguish between two gravity models rather than to observe detailed peak structures, we can sacrifice some frequency resolution by reducing the measurement time to, for example, $T_{\rm obs}=40$\,sec. This approach significantly reduces measurement-induced shot noise, enabling model discrimination with a false alarm rate of just 2\% after only $10^3$ repeated experimental trials—representing an improvement in experimental efficiency, though is still cost-prohibitive. 

The cost-prohibition arises because the WV spectrum analysis only extracts information from the anti-diagonal elements of the full covariance matrix, thereby under-utilizing the complete statistical information available in the measurement data. To address this inefficiency, we developed a statistical inference framework that leverages the entire covariance matrix structure, which significantly reduces experimental requirements while maintaining high discriminative power.

Our statistical inference method treats the complete time-series data as a high-dimensional Gaussian random variable, enabling us to construct and analyze the full covariance matrix rather than extracting only partial spectral information. By comparing the log-likelihood ratios of Mock data against theoretical predictions from both gravity models, we can achieve reliable model discrimination with significantly fewer experimental repetitions. Remarkably, our analysis demonstrates that with an initial squeezed thermal state of only 1.5\,dB and just 10 experimental trials of 40\,sec each, we can distinguish between CCSN and QG theories while maintaining false alarm rates below 1\%.

The experimental realization of our proposal presents several key challenges that must be addressed: First, achieving a millihertz-frequency oscillator with an ultra-high Q-factor of $10^7$ remains technically demanding. Current state-of-the-art systems, as demonstrated in Ref.~\cite{PhysRevLett.124.221102}, have successfully realized mechanical oscillators operating at $2.2$\,Hz with Q-factors of $10^6$ at room temperature. Second, preparing the test mass in the required squeezed thermal state poses significant experimental difficulties. While higher squeezing levels generally enhance the distinguishability between competing theoretical models, current experimental capabilities reached approximately 4.5\,dB of squeezing for Hz-frequency mechanical oscillators~\cite{santiagocondori2023}.  Encouragingly, our theoretical analysis demonstrates that these stringent requirements can be substantially relaxed: for millihertz mechanical oscillators, modest squeezing levels as low as 1.5\,dB may prove sufficient for model discrimination, though this relaxation would require additional experimental repetitions to maintain adequate statistical significance. Additionally, considering that low-frequency oscillators are inherently more susceptible to low-frequency environmental perturbations such as seismic noise, advanced vibration isolation technologies become essential. Several promising solutions can be adopted to address these challenges, including active isolation systems~\cite{app10207342,fernándezgaliana2019}, passive isolation techniques~\cite{Hermsdorf2018HighPP}, or sophisticated hybrid isolation approaches~\cite{Lee2017}.

In the future, non-stationary evolution in mutual gravity scenarios may exhibit novel phenomena; for instance, Ref.~\cite{Liu2025} demonstrates that "apparent entanglement" can be established between two outgoing light fields via mutual SN gravity, whereas classical gravity, as a LOCC-type interaction, cannot entangle the two test masses. By analyzing the non-stationary evolution of the entire system, we can illuminate how entanglement is generated by mutual SN gravity and clarify its distinctions from entanglement generation in quantum gravity.

\appendix
\section{The spectrum of outgoing field for CCSN theroy under quantum thermal noise prescription}\label{Appendix A}

In main text, we analyzed the scenario where SN theory is applied to examine the spectrum of output light with thermal noise described as classical Brownian motion. In this appendix, we treat thermal noise as originating from a quantum thermal reservoir. While the spectrum of the output light maintains the same overall structure as in the classical thermal noise case, the relative difference between SN theory and QG is significantly enhanced.

Alternatively, the thermal environment can be modeled as a quantum reservoir~\cite{Hopkins2003}. Since we do not directly measure the thermal noise, tracing out the reservoir degrees of freedom leads to the following SME:
\begin{equation}\label{eq:SME for quantum thermal noise}
\begin{split}
        \frac{d\hat{\rho} }{dt}=& -\frac{i}{\hbar}[\hat{H}_{\rm{0}},\hat{\rho}]+\frac{\alpha}{\sqrt{2}}\{\hat{x}-\langle\hat{x}\rangle,\hat{\rho}\}\frac{dW}{dt}-\frac{\alpha^2}{4}[\hat{x},[\hat{x},\hat{\rho}]]\\
        &-\frac{i\gamma_m}{2\hbar}[\hat{x},\{\hat{p},\hat{\rho}\}]-\frac{M\omega_q\gamma_m}{2\hbar}\coth{\frac{\hbar\omega_q}{2k_bT}}[\hat{x},[\hat{x},\hat{\rho}]].
\end{split}
\end{equation}

From Eq.~\eqref{eq:SME for quantum thermal noise}, we can derive the evolution of the conditional mean. This naturally incorporates the mechanical damping rate:
\setlength{\jot}{10pt} 
\begin{align}
    \frac{d\langle \hat{x}\rangle_c}{dt} &= \frac{\langle \hat{p}\rangle_c}{M} + \sqrt{2}\alpha V_{xx}(t)\frac{dW}{dt}, \label{eq:mean-of-x} \\
    \frac{d\langle \hat{p}\rangle_c}{dt} &= -M \omega_m^2 \langle \hat{x}\rangle_c - \gamma_m\langle \hat{p} \rangle_c 
    + \sqrt{2}\alpha V_{xp}(t)\frac{dW}{dt}. \label{eq:mean-of-y}
\end{align}
Similarly, we can also obtain the evolution equation regarding the conditional second-order moment:
\setlength{\jot}{10pt} 
\begin{align}
    \dot{V}_{xx} &= \frac{2}{M}V_{xp} - 2\alpha^2V_{xx}^2, \label{eq:Vxx} \\
    \dot{V}_{xp} &= \frac{V_{pp}}{M} - M\omega_q^2V_{xx} - 2\alpha^2V_{xp}V_{xx} - \gamma_mV_{xp}, \label{eq:Vxp} \\
    \dot{V}_{pp} &= -2M\omega_q^2V_{xp} - 2\gamma_mV_{pp} - 2\alpha^2V_{xp}^2 + \frac{\hbar^2\alpha^2}{2} \notag \\
    &\quad + M\hbar\omega_q\gamma_m\coth[\hbar\omega_q/(2k_{b}T)]. \label{eq:Vpp}
\end{align}
The additional terms introduced by the quantum thermal noise cause the second-order moments to converge to a thermal equilibrium state.

In this case, when the system is in a steady state, the spectrum of outgoing light $\tilde{y}(t)$ is:
\begin{equation}
\begin{split}
        S_{\tilde{y}\tilde{y}}^{(\rm{SN})}(\Omega)=&\frac{\omega_q^4}{(\Omega^2-\omega_m^2)^2+\gamma_m^2\Omega^2}
        \Bigg[\Lambda_q^4+\frac{2\Lambda_q^2}{Q_q}\coth{\frac{\hbar\omega_q}{2k_b T}}\\
        &-2\frac{\omega_{\rm{SN}}^2}{\omega_q^2}\left(\sqrt{1+\Lambda_q^4+\frac{2\Lambda_q^2}{Q_q}\coth{\frac{\hbar\omega_q}{2k_b T}}}-1\right)\Bigg]+1,
\end{split}
\end{equation}
where the first term and the second term represent the response of the light and quantum thermal noise; the third term is the SN gravity effect in causal-condition prescription; the last term is the shot noise. 

As similar, we can use the logarithmic ratio $\mathcal{S}(\Omega)$ to evalute the relative difference of the two models. We find the $\mathcal{S}(\omega_m)$ can reach its maximum value:
\begin{equation}
    -10\log_{10}\left[1-\frac{2\omega_{\rm{SN}}^2}{2\gamma_m\omega_m + 2\omega_q^2}\right],
\end{equation}
when the $\Lambda_q = \sqrt{\hbar\omega_m/(2k_B T)}$. Under the same system parameters, the logarithmic ratio $\mathcal{S}(\omega_m)$ can reach 26. Although theoretical differences exist between the two models, achieving such large differences experimentally is challenging. At currently achievable experimental temperatures (approximately 1\,K to 1\,mK), the optimal cavity power is so weak that the photon number in the cavity is much less than 1. However, even under non-optimal conditions, the quantum thermal noise prescription still exhibits significantly greater model discrimination compared to the classical thermal noise prescription. For example, at a temperature of 1\,K and cavity power of $1\,\mu$W, the logarithmic ratio $\mathcal{S}(\omega_m)$ can reach 8.

\section{Peak conditions for damped oscillation spectrum}\label{Appendix B}
In the main text, we derived the oscillatory form of $h_1$ near the equilibrium point. Here, we present a simple proof demonstrating that under strong measurement conditions, the oscillatory behavior of $h_1$ will be suppressed due to rapid decay.

Consider the simple form of the damped oscillation function:
\begin{equation}
    f(t)=e^{-\lambda t}\sin{\omega t}H(t),
\end{equation}
where $H(t)$ is the Heaviside step function. It can be proved that the condition $\lambda\ge\omega$ is satisfied, the spectrum of $f(t)$ has no peak.

The square of spectrum of $f(t)$ can be calculated by using the Fourier transform:
\begin{equation}
    \begin{split}
        |F(\Omega)|^2=&|\mathcal{F}[f(t)]|^2=\left|\frac{\omega}{\omega ^2+(\lambda -i \Omega )^2}\right|^2\\
        =&\frac{\omega^2}{\left(\lambda ^2+\omega ^2\right)^2+2 \Omega ^2 (\lambda -\omega ) (\lambda +\omega )+\Omega ^4}
    \end{split}
\end{equation}
Since $|F(\Omega)|^2$ has the same monotonicity as $|F(\Omega)|$, we only need to analyze the monotonicity of $|F(\Omega)|^2$. The numerator of $|F(\Omega)|^2$ is a constant, which does not influence the monotonicity of $|F(\Omega)|^2$. Therefore, the existence of a peak structure in the spectrum is determined by the denominator.

Let the first derivative of the denominator to be zero:
\begin{equation}
    4 \Omega  (\lambda -\omega ) (\lambda +\omega )+4 \Omega ^3=0,
\end{equation}
we can obtain three roots:
\begin{equation}
    \Omega_1=0,\ \Omega_2=\sqrt{\omega^2-\lambda^2},\ \Omega_3=-\sqrt{\omega^2-\lambda^2}.
\end{equation}
If $\lambda\ge\omega$, there only one real root $\Omega_1=0$, we can easily find the derivative of the denominator is always positive for $\Omega>0$. That indicates that the denominator is a monotonically increasing function with respect to $\Omega$. As a result, the spectrum of the damped oscillation function does not exhibit a peak.

For the case $0<\lambda<\omega$, the first derivative is negative in the range $0<\Omega<\sqrt{\omega^2-\lambda^2}$, and positive for $\Omega>\sqrt{\omega^2-\lambda^2}$. This implies that the denominator attains a minimum at $\Omega=\sqrt{\omega^2-\lambda^2}$.  Consequently, the spectrum of the damped oscillation function exhibits a peak at this frequency.

\section{The relationship between the Fourier transform of the variable limit integral and the Fourier transform of the integrand}
\label{Appendix C}

In the process of calculating the WV spectrum, we encounter a variable integral of the form $\int_{0}^{t-\frac{|\tau|}{2}} ds \, f(s)$. We observe that there exists a specific relationship between the oscillation frequency of this variable integral and that of its integrand. This relationship can be rigorously established through Fourier analysis, as demonstrated in the following proof.

Let $f(s)$ be an arbitrary function. Its Fourier transform, denoted by $\tilde{f}(\Omega)$, is defined as:
\begin{equation}
    \tilde{f}(\Omega) = \int_{-\infty}^{+\infty} ds \, f(s) e^{-i\Omega s}.
\end{equation}
We consider the following integral as a function of $\tau$:
\begin{equation}\label{eq:integral_function}
    I(t, \tau) = \int_{0}^{t-\frac{|\tau|}{2}} ds \, f(s),
\end{equation}
where $t$ is a fixed parameter greater than 0. This integral is defined for $\tau \in [-2t, 2t]$. For $|\tau| > 2t$, we define $I(t, \tau) = 0$.

We aim to obtain the relationship of the Fourier transform of $I(t, \tau)$ between $F(\Omega)$, we need to do the Fourier transform of $I(t, \tau)$, which can be expressed as:
\begin{equation}\label{eq:Fourier_transform_of_I}
\begin{split}
    &\int_{-\infty}^{+\infty}d\tau\ e^{-i\Omega\tau} \left[\int_{0}^{t-\frac{|\tau|}{2}}ds\ f(s)\right].
\end{split}
\end{equation}

First, performing the inverse Fourier transform on $f(s)$ and exchanging the order of integration, the Eq.~\eqref{eq:Fourier_transform_of_I} can be rewritten as:
\begin{equation}
    \begin{split}
        &\int_{-2t}^{2t}d\tau\ e^{-i\Omega\tau}\int_{0}^{t-\frac{|\tau|}{2}}ds \left[\int_{-\infty}^{+\infty}\frac{d\Omega'}{2\pi} \tilde{f}(\Omega')e^{i\Omega's}\right]\\
        =&\int_{-\infty}^{+\infty}\frac{d\Omega'}{2\pi} \tilde{f}(\Omega')\left[\int_{-2t}^{2t}d\tau\int_{0}^{t-\frac{|\tau|}{2}}ds\ e^{i\Omega's}e^{-i\Omega\tau}\right]\\
        =&\int_{-\infty}^{+\infty}\frac{d\Omega'}{2\pi} \tilde{f}(\Omega')\frac{4 e^{i\Omega't}\Omega-4\Omega\cos{2\Omega t}-2i\Omega'\sin{2\Omega t}}{\Omega(4\Omega^2-\Omega'^2)}.
    \end{split}
\end{equation}
Next, performing the Fourier transform on $F(\Omega')$ again and exchanging the order of integration, we have the following relationship:
\begin{equation}\label{eq:C5}
    \begin{split}
        &\int_{-\infty}^{+\infty}\frac{d\Omega'}{2\pi}\frac{4 e^{i\Omega't}\Omega-4\Omega\cos{2\Omega t}-2i\Omega'\sin{2\Omega t}}{\Omega(4\Omega^2-\Omega'^2)}\\
        &\quad\quad\quad\quad\ \times\left[\int_{-\infty}^{+\infty}dt' f(t')e^{-i\Omega't'}\right]\\
        =&\int_{-\infty}^{+\infty}dt' f(t')G(t,t'),
\end{split}
\end{equation}
where the $G(t,t')$ is:
\begin{equation}
\begin{split}
    &\int_{-\infty}^{+\infty}\frac{d\Omega'}{2\pi}e^{-i\Omega't'}\frac{4 e^{i\Omega't}\Omega-4\Omega\cos{2\Omega t}-2i\Omega'\sin{2\Omega t}}{\Omega(4\Omega^2-\Omega'^2)}\\
    =&\frac{\sin{2\Omega(t-t')}-\sin{2\Omega(t+t')}}{\Omega}.
\end{split}
\end{equation}
Then, the Eq.~\eqref{eq:C5} becomes:
\begin{equation}
\begin{split}
        &\int_{-\infty}^{+\infty}dt' f(t')\frac{\sin{2\Omega(t-t')}-\sin{2\Omega(t+t')}}{\Omega}\\
        =&-\frac{2\cos{2\Omega t}}{\Omega}\int_{-\infty}^{+\infty}dt' f(t')\sin{2\Omega t'}\\
        =&\frac{2\cos{2\Omega t}}{\Omega}\rm{Im}[\tilde{f}(2\Omega)].
    \end{split}
\end{equation}

\section{The error rate in the quantum thermal noise case}

\begin{figure}[h]
    \centering
        \includegraphics[scale=0.43]{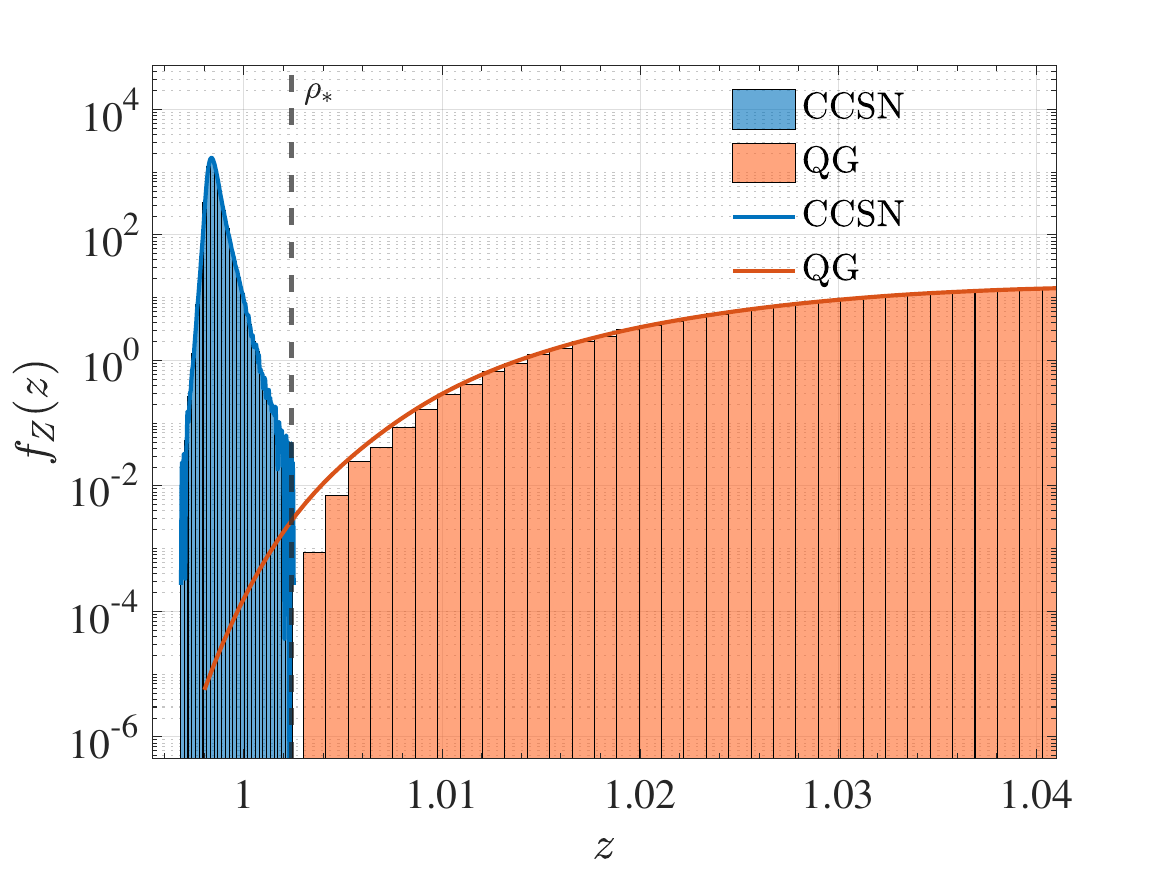}
        \caption{Probability density distributions of the log-likelihood ratio $Z$ under the quantum thermal noise prescription, calculated using the same 4.5 dB squeezed thermal state. The error rate is $\mathcal{F}=2.3\times10^{-6}$.}
        \label{fig:PDF of z quantum noise}
\end{figure}

In this section, we will discuss the impact of the quantum thermal noise prescription on non-stationary evolution and analyze the false alarm rate for distinguishing between the two gravitational models using statistical inference methods.

Following the same approach as in the classical thermal noise case, the conditional displacement can be obtained by solving Eq.~\eqref{eq:mean-of-x} and Eq.~\eqref{eq:mean-of-y}:
\begin{equation}
    \langle \hat{x}(t)\rangle_c = e^{-\frac{\gamma_m}{2}t}\left[x^{(0)}(t)+x^m(t)\right],
\end{equation}
where $x^{(0)}(t)$ and $x^m(t)$ are identical to those in the classical thermal noise case. The key difference is that the thermal-noise-induced displacement $x^{\rm th}(t)$ is absorbed into the conditional second-order moments. The comparison between these two prescriptions reveals fundamental differences in how thermal noise affects the SN signal. Under the classical thermal noise prescription, thermal fluctuations do not directly modify the strength of the SN signature itself, but instead contribute additional noise to the output optical signal. In contrast, the quantum thermal noise prescription directly influences the evolution dynamics of the conditional second-order moments, thereby modifying the very nature of the signal that carries the SN information.

To quantitatively assess the impact of thermal noise on the signal under these two prescriptions, we can provide a simplified estimation. Under the quantum thermal noise prescription, the signal $\alpha x^{(m)}(t)$ can be conceptually separated into two components: the SN signal and quantum thermal fluctuations that suppress the SN signal. The intensity of this quantum thermal suppression is proportional to $\Lambda_s/\sqrt\omega_q$. In contrast, classical thermal noise $\alpha x^{\rm th}(t)$ is directly added to the signal with an intensity proportional to $\Lambda_s/\sqrt{\omega_m}$. Consequently, under conditions where the conditional second-order moments maintain their oscillatory behavior without rapid convergence, the quantum thermal noise prescription consistently enhances the distinguishability between the two gravitational models by providing a more favorable signal-to-noise ratio. This theoretical prediction is confirmed by our statistical inference analysis. As demonstrated in Fig.~\ref{fig:PDF of z quantum noise}, under identical experimental parameters and using the same 4.5 dB squeezed thermal state, the quantum thermal noise prescription achieves a remarkably low false alarm rate of $\mathcal{F}=2.3\times10^{-6}$ after averaging over ten experimental trials.

\acknowledgements
Y.L. and Y.M. want to thank Dr Tianliang Yan, Professors Yanbei Chen, Haixing Miao and Denis Marytnov for discussion on this topic. Y.M. is supported by the National Key R\&D Program of China “Gravitational Wave Detection"\,(Grant No.2023YFC2205801), National Natural Science Foundation of China under Grant No.12474481, No.12441503, and the start-up funding provided by Huazhong University of Science and Technology.

\bibliographystyle{unsrt}
\bibliography{causal-conditional.bib}

\end{document}